\documentclass[useAMS,fleqn,usenatbib]{mnras}

\usepackage{graphicx}
\usepackage{mathtools}
\usepackage[usenames,dvipsnames]{color}
\usepackage{amsmath,amssymb,latexsym}

\usepackage[T1]{fontenc} 
\usepackage{ae,aecompl}

\newcommand{\be}{\begin{equation}}
\newcommand{\ee}{\end{equation}}
\newcommand{\bea}{\begin{eqnarray}}
\newcommand{\nn}{\nonumber}
\newcommand{\eea}{\end{eqnarray}}

\newcommand{\p}{\partial}

\title[EoS-independent neutron star properties]{Unified description of astrophysical properties of neutron stars independent of the equation of state}
\author[G.~Pappas]{George Pappas$^{1}$\thanks{E-mail:
georgiospappasgr@gmail.com}\\ 
$^{1}$School of Mathematical Sciences, The University of Nottingham,
University Park, Nottingham NG7 2RD, UK}

\date{Accepted 2015 September 22. Received 2015 September 17; in original form 2015 July 3}

\pubyear{2015}

\begin{document}
\label{firstpage}
\pagerange{\pageref{firstpage}--\pageref{lastpage}}
\maketitle

\begin{abstract}
In recent years, a lot of work was done that has revealed some very interesting properties of neutron stars. One can relate the first few multipole moments of a neutron star, or quantities that can be derived from them, with relations that are independent of the equation of state (EoS). This is a very significant result that has great implications for the description of neutron stars and in particular for the description of the spacetime around them. Additionally, it was recently shown that there is a four parameter analytic spacetime, known as the two-soliton spacetime, which can accurately capture the properties of the geometry around neutron stars. This allows for the possibility of describing in a unified formalism the astrophysically relevant properties of the spacetime around a neutron star independently of the particulars of the EoS for the matter of the star. More precisely, the description of these astrophysical properties is done using an EoS omniscient spacetime that can describe the exterior of any neutron star. In the present work we investigate properties such as the location of the innermost stable circular orbit $R_{ISCO}$ (or the surface of the star when the latter overcomes the former), the various frequencies of perturbed circular equatorial geodesics, the efficiency of an accretion disc, its temperature distribution, and other properties associated with the emitted radiation from the disc, in a way that holds for all possible choices of a realistic EoS for the neutron star. Furthermore, we provide proof of principle that if one were to measure the right combinations of pairs of these properties, with the additional knowledge of the mass of the neutron star, one could determine the EoS of the star.      
\end{abstract}


\section{Introduction}

Low mass X-ray binaries (LMXBs) and the neutron stars (NSs) that reside in them are sources of some of the most interesting astrophysical phenomena. They also provide the opportunity to study the most extreme physics in nature. The processes involved in these astrophysical objects as far as the NSs are concerned, are related both to strong gravity effects and effects associated with the properties of matter in densities that are above the nuclear density.  
Therefore these systems could be used as tools for broadening both our understanding of gravity, by testing the predictions of the established theory of General Relativity (GR), and our understanding of the equation of state (EoS) of matter for densities as high as the ones that can be found at the centres of NSs. 

Thus far, the uncertainty in the properties of the EoS on the one hand and the fact that there are a lot of physical processes that are taking place in the interior and around NSs, has made them ``dirty" systems for extracting clean signals for gravity. This picture partially changed recently with the discovery of the universal behaviour that NSs exhibit with respect to their structure, i.e., behaviour that is independent of the EoS used for the nuclear matter inside the NS. It was shown initially by \cite{YY2013Sci,YY2013PhRvD} that for slowly rotating NSs there exist EoS-independent relations between the normalized moment of inertia $\bar{I}\equiv I/M^3$ (where $M$ is the mass), the normalized or reduced quadrupole $\bar{Q}\equiv - Q/(M^3j^2)$\footnote{This quantity will also be referred to as quadrupolar deformability.} (where $j\equiv J/M^2$ is the spin parameter and $J$ is the angular momentum), and the normalized Love number $\bar{\lambda}\equiv \lambda/M^5$. \cite{Pappas:2013naa} generalised the universality of the $\bar{I}-\bar{Q}$ relations for arbitrarily rotating NSs\footnote{Almost simultaneously \cite{Chakrabarti2014PRL} came out with the same results and some further extensions.} and further showed that a more fundamental universality holds, related to the spacetime properties, that connects the first four multipole moments in an EoS-independent way. In their work, \cite{Pappas:2013naa} additionally showed that if one were to measure the first three multipole moments of the spacetime around a NS, then one could distinguish an EoS for the star from the known realistic EoSs. The universal relations for the multipole moments in GR have been extended up to the mass hexadecapole, $M_4$, which is the fifth non-zero moment, and have been shown to describe apart from NSs, quark stars as well (see the work by \cite{YagietalM4} and also see the work by \cite{Stein2014ApJ} for a generalised Newtonian treatment).     

For the LMXBs, there are several properties (or observables in some cases) that can be associated to the properties of the spacetime and more specifically to the properties of geodesics around the compact object that is part of such a system. 

Quasi-periodic oscillations (QPOs) of the X-ray flux observed from LMXBs is one example (for a review see \cite{lamb,derKlis}), where one of the proposed mechanisms for producing this behaviour is the relativistic precession of geodesics (see \cite{stella}). This mechanism for explaining QPOs is not the only one that has been proposed, but it is the most straightforwardly related to properties of geodesic motion. In the context of the relativistic precession model, one assumes that fluid elements of an accretion disc follow geodesic or almost geodesic orbits which are almost circular and have some inclination with respect to the equatorial plane (an assumption that can be considered valid for radiatively efficient accretion discs). General relativity predicts that these orbits will generally display a precession of the periastron of the orbit and a precession of the orbital plane, or of the nodes of that orbit. Therefore, there are several frequencies associated to these orbits, the orbital frequency of the circular motion $\Omega$, the radial oscillation of the fluid element along its orbit that has frequency $\kappa_{\rho}$ and the corresponding precession frequency $\Omega_{\rho}\equiv\Omega-\kappa_{\rho}$ (periastron precession), and finally the vertical oscillation of the fluid element with respect to the equatorial plane which has frequency $\kappa_z$ and the corresponding precession frequency $\Omega_z\equiv\Omega-\kappa_z$ (nodal precession). Here, the term ``frequency" refers to the angular velocity that is equal to $2 \pi \nu$, where $\nu$ is the frequency expressed as cycles in the unit of time. 

Another property that can be associated to the geometry of the spacetime around a compact object that is surrounded by an accretion disc, is the total luminosity radiated from the disc. This is a property that is associated to the efficiency $\eta=1-\tilde{E}_{\rm ISCO}$ of the disc, where $\tilde{E}_{\rm ISCO}$ is the energy per unit mass of a particle that has a circular orbit at the location of the innermost stable circular orbit (ISCO), which is essentially the total energy that a particle of unit mass loses as it moves from infinity up to the ISCO. If one assumes a thin radiatively efficient accretion disc, then a fluid element of unit mass in the disc will have radiated by the time it plunges from the ISCO to the compact object, a total amount of energy that is equal to $\eta$ times its mass.

Along the same lines, the temperature distribution of a thin radiatively efficient disc can be associated to the variation of $\tilde{E}$ of circular geodesics with the radial distance from the central object. Therefore, the emmited spectrum of the disc can be associated to the geometry of the spacetime. This type of properties of an accretion disc and their connection to the background geometry have been used in the past to determine the rotation of black holes (see for example \cite{narayan} and references therein), where a technique called, continuum-fitting method, is applied in order to determine the spin of the black hole from the luminosity of the disc.

\subsection{Executive summary}

This work is relatively lengthy and there are several different ideas that are being discussed. Therefore, in order to help the reader, we provide here a brief summary of the topics discussed. 

In what follows we give a universal, in the sense of EoS-independent or more precisely not EoS-specific, description of the spacetime around NSs and consequently a universal and EoS-independent description of various astrophysical phenomena that take place around these NSs. To have this EoS-independent description of the spacetime we use the analytic two-soliton spacetime, that has been shown to be an accurate representation of the spacetime around NSs (see work by \cite{twosoliton}), in conjunction with the universal three-hair relations found for NSs in GR. This results in a description of the spacetime in terms of the mass and two more parameters, the spin parameter $j$ and the quadrupolar deformability parameter $\alpha\equiv\bar{Q}$. The mass essentially introduces only a scale and can therefore be removed from the various observables by applying the appropriate reduction. Therefore we end up with a parametric description of the various mass-reduced observables in terms of only two parameters, $j$ and $\alpha$.  
The universal spacetime description is supplemented by a discussion of two more approximately EoS-independent relations that give the mass-reduced  equatorial radius and the mass-reduced rotation frequency of NSs, in terms of the parameters $j$ and $\alpha$. 

The aforementioned observables have been separated into two categories, the observables that are related to the characteristics of geodesic motion on and around the equatorial plane, which are mainly the various frequencies of the orbital motion, and the observables that are related to the radiation emitted by an accretion disc that forms around a NS that accretes matter. 

The observables discussed here, that belong to the first category, are the orbital and the nodal precession frequency of circular orbits at the location of the innermost available orbit around the NS. That orbit can either be the ISCO or the circular orbit that is just outside the stellar surface. The closest orbit to the star has been chosen because it is a characteristic orbit around the star. Another characteristic orbit that some stars have and belongs to the first category, is the orbit where the nodal precession becomes zero. This is a novel property that NS spacetimes have that rotating black hole (Kerr) spacetimes don't exhibit. At that location the observables that we consider are the orbital frequency and the periastron precession frequency.      

For the second category of observables, we consider the accretion disc efficiency, the temperature distribution of an accretion disc and in particular the maximum temperature, the photon energy where the spectrum has the maximum integrated luminosity, i.e., the photon energy where there is the maximum energy output from the disc, and finally the maximum integrated luminosity itself. The latter quantity turns out to be equivalent to the accretion efficiency, as one would expect. 

Finally we briefly discuss how could one use the various observables to perform consistency checks of the various proposed phenomenological models that relate astrophysical observables, such as QPOs, to properties of the geodesics around NSs, or how well accretion discs for example are described by a thin radiatively efficient disc model. Furthermore we discuss how could one measure the first three multipole moments of a NS by using observations of different observables. The latter application could be used to constrain and maybe eventually determine the EoS of nuclear matter at supra-nuclear densities. 

We should emphasize again that the main underlying assumption in our analysis is that the motion of particles or fluid elements is geodesic. This implies some restrictions for the systems for which this analysis is applicable. One such restriction is that the central object should have a low magnetic field, because in a different case, the large scale magnetic field could modify the orbital motion of fluid elements in the inner part of the disc, that would then deviate from being geodesic. Another related restriction would be that the disc is in a relatively quiet state with a moderate accretion rate, so that the steady state thin accretion disc model is applicable. These requirements impose selection criteria for the LMXB systems that one could use. 

But one could go beyond the analysis presented here, by assuming a different model for accretion or different mechanisms for the QPOs. Even in that case, the main idea behind this analysis would remain the same, i.e., that one could use an analytic background spacetime that can be parameterized by only two parameters in order to express astrophysical observables in terms of only these two spacetime parameters. The difference in that case though would be that in contrast to the relativistic precession model for QPOs and the radiatively efficient thin disc model for the accretion, one would have to include some more physical parameters that would depend on the particular model.

\subsection{Plan of the paper}

The plan of the paper is as follows, in section \ref{sec:2} we briefly present the analytic spacetime that describes the exterior of NSs and discuss how we implement the universal relations between the moments to arrive to a universal NS spacetime. In section \ref{sec:observables} we first discuss the observables related to geodesic motion and then we discuss the observables related to the radiation emitted by an accretion disc around a NS. In section \ref{sec:4} we discuss how could one apply the results of section \ref{sec:observables} to testing the various models for the astrophysical phenomena observed around NSs and how one could measure the first three multipole moments of the NS. We end with the conclusions in section \ref{sec:conclusions}, while there is some further discussion in the appendix on an approximately universal relation for the equatorial radius of NSs (appendix \ref{radii}), on an approximately universal relation for the rotation frequency of NSs (appendix \ref{NSrotation}), and on the properties of the nodal precession frequency for NSs (appendix \ref{app:Nodal}). For the calculation we use geometric units, where $G=c=1$ and the mass is given in $km$, unless some other unit is specified (usually for frequencies, temperature and energy).

\section{The spacetime around NSs}
\label{sec:2}

Determining the spacetime around rotating compact fluid configurations is a difficult problem that can be solved analytically only in the slow rotation limit (although one still has to integrate numerically the interior of the star). If one were to tackle the problem of rapidly rotating configurations one would have to resort to a fully numerical solution. The slow rotation solution is given by the \cite{HT} approach and there has been a lot of work extending this formalism up to fourth order in rotation (see work by \cite{berti-white} and \cite{YagietalM4}). Similarly there has been a lot of work on numerical algorithms for solving the full Einstein field equations for axisymmetric spacetimes around rotating fluid configurations (for example see \cite{Sterg} and for more details the review by \cite{Lrr}). Additionally to these two approaches there have been investigations of analytic axisymmetric spacetimes that can match and accurately describe the exterior of rotating fluid configurations (see for example work by \cite{Stute,berti-stergioulas,Pappas2,Teich,Pachon,twosoliton}). In particular, in \cite{Pappas2,twosoliton} it was shown that the two-soliton analytic solution of \cite{twosoliton2} can describe accurately the exterior spacetime of a NS for arbitrary rotation, rapid or slow. For our purposes here, we will use for the spacetime exterior to NSs the two-soliton analytic spacetime. 

The two-soliton solution is a four parameter solution of the vacuum Einstein's field equations that is generated using the Ernst potential and the algorithm developed by \cite{sib1,SibManko,manko2,twosoliton2}.     

The vacuum region of a stationary and axially symmetric spacetime in GR
can be described by the line element first used by \cite{Papapetrou}
\be \label{Pap} ds^2=-f\left(dt-\omega d\phi\right)^2+
      f^{-1}\left[ e^{2\gamma} \left( d\rho^2+dz^2 \right)+
      \rho^2 d\phi^2 \right],
\ee
where $f,\;\omega,$ and $\gamma$ are functions of the
Weyl-Papapetrou coordinates ($\rho,z$). By introducing the complex
potential $\mathcal{E}(\rho,z)=f(\rho,z)+\imath\psi(\rho,z)$,
\cite{ernst1} reformulated the Einstein field equations in the form of the complex equation
\be \label{ErnstE}
(Re(\mathcal{E}))\nabla^2\mathcal{E}=\nabla\mathcal{E}\cdot\nabla\mathcal{E},
\ee
where $\nabla$ and $\nabla^2$ are respectively the gradient and the Laplacian in flat cylindrical coordinates $(\rho,z,\phi)$.

One can generate a solution of the Ernst equation by making a choice for the Ernst potential along
the axis of symmetry of the spacetime, in the form of a rational function
\be
\mathcal{E}(\rho=0,z)=e(z)=\frac{P(z)}{R(z)},
\label{Eernst}
\ee
where $P(z), R(z)$ are polynomials of $z$ of order $n$ with complex coefficients in general. 

The vacuum two-soliton solution (proposed by \cite{twosoliton2}) is a special case of the previous general axisymmetric solution that
is obtained from the ansatz 
\be
\label{2soliton}
e(z)=\frac{(z-M-ia)(z+ib)-k}{(z+M-ia)(z+ib)-k},
\ee
where all the parameters $M, a, k, b$ are real (for more details on the algorithm for generating solutions and for the two-soliton spacetime in particular see \cite{twosoliton2,twosoliton}).

The first few multipole moments of the two-soliton spacetime are expressed in terms of the four parameters of the Ernst potential as,
\bea
\label{moments2}
M_0&=&M,\quad  M_2=-(a^2-k)M,\nn\\
 M_4&=&\left[ a^4 - (3a^2 -2ab + b^2)k + k^2 - \frac{1}{7}kM^2\right] M\nn\\
J_1&=& aM ,\quad J_3=- [a^3 -(2a - b)k]M,
\eea
where $M_0=M$ is the mass, $M_2=Q$ is the quadrupole moment, $M_4$ is the mass hexadecapole, $J_1=J$ is the angular momentum, and $J_3$ is the spin octupole moment. 

It was shown by \cite{twosoliton} that if one were to chose the parameters $M, a, k, b$ in such a way so that the first four non-zero moments of the two-soliton spacetime were equal to the moments of the numerical spacetime (calculated using the prescriptions presented in \cite{pappas-apostolatos} and further elaborated in \cite{YagietalM4}), then the analytic solution would be an accurate match for the numerical solution and that it would capture all the properties of the geodesics that are relevant to the astrophysical processes that take place in LMXBs.

Furthermore, \cite{Pappas:2013naa} have found that for NSs (and this has been extended to quark stars as well by \cite{YagietalM4}) the first multipole moments are not all independent between them. In particular one can express the first moments that are higher than the quadrupole in terms of the quadrupole. Specifically, if we define the reduced moments as
\be \bar{M}_n=\frac{M_n}{j^nM^{n+1}}, \; \bar{J}_n=\frac{J_n}{j^nM^{n+1}}, \ee
where $j$ is the spin parameter defined as $J/M^2$, then the spin octupole and the mass hexadecapole of a NS will be related to the quadrupole by relations of the form 
\be  y=A+B_1 x^{\nu_1}+B_2 x^{\nu_2} ,\ee 
where $y$ can be either $\sqrt[3]{-\bar{J}_3}$ or $\sqrt[4]{\bar{M}_4}$ and $x$ is $\sqrt{-\bar{M}_2}$. Therefore the first higher moments of a NS spacetime can be expressed in terms of only three parameters, the mass $M$, the angular momentum $J$, and the quadrupole $M_2=Q$. 

Additionally, as it was shown by \cite{Pappas:2013naa} and \cite{YagietalM4}, these relations between the moments are independent of the EoS which means that if one were to use these expressions to produce a two-soliton spacetime, then one could have an EoS independent description of the spacetime around a NS. To clarify the last statement, using the universal relations between the moments, one could construct a spacetime metric parameterized in such a way, so as to be suitable to describe any NS having any one of the realistic EoSs that we have at our disposal. This EoS-independent (or -universal) description of the NS spacetime is what we aim to have here. 

Since for the construction of the spacetime we have only four parameters which are fixed by the first four moments, $M,J,M_2,$ and $J_3$, we will only need the relation between the quadrupole and $J_3$ given by \cite{Pappas:2013naa}, i.e., 
\be y=-0.36+1.48\, x^{0.65},\ee
where $y=\sqrt[3]{-\bar{J}_3}$ and $x=\sqrt{-\bar{M}_2}$. 
For the description of the spacetime and the various properties of the geodesics that we will calculate, we will use as a first parameter the dimensional mass $M$, expressed in units of km, and two additional dimensionless parameters which will be the spin parameter $j=J/M^2$ and the reduced quadrupole $\alpha=-M_2/(j^2M^3)$. 

%
%

\section{Astrophysical properties and observables}
\label{sec:observables}

As it was elaborated in the previous section, we now have a three parameter description of the spacetime around a NS, which is independent of a particular choice for the EoS of the matter inside the NS. The three parameters that enter the description are the mass $M$ that gives the scale to all the quantities and the two dimensionless parameters $j$, which is the spin parameter and is a measure of the rotation of the NS, and $\alpha$, which is the reduced quadrupole and is a measure of the deformability due to rotation. For this reason we will also call the parameter $\alpha$ the quadrupolar deformability parameter. Since the mass $M$ is a scale of the spacetime and its properties, by appropriately reducing or normalizing the various quantities we can end up with a two-parameter description of all the astrophysical observables, as we will see in what follows.

\subsection{Equatorial geodesics in an axisymmetric spacetime}

We start with a brief discussion of geodesics in an axisymmetric spacetime. An axisymmetric spacetime given by the line element in eq~\eqref{Pap} admits two Killing fields which define the symmetries with respect to translations in time and rotations with respect to an axis. One can take advantage of these symmetries and the corresponding integrals of motion to study geodesics in the spacetime. 
The first integral of motion is the energy $E$, given as
\be E=-p_a\xi^a=-p_t
   =\mu\left(-g_{tt}\frac{dt}{d\tau}-g_{t\phi}\frac{d\phi}{d\tau}\right),\ee
where $t$ is the coordinate time and $\tau$ is proper time. 
The second integral of motion is the angular momentum $L$ in the direction of the symmetry axis, given as 
\be L=p_a\eta^a=p_{\phi}
   =\mu\left(g_{t\phi}\frac{dt}{d\tau}+g_{\phi\phi}\frac{d\phi}{d\tau}\right)\ee 
In addition  to the two previous equations we have from the normalization of the four-velocity the equation 

\bea -1&=&g_{tt}\left(\frac{dt}{d\tau}\right)^2+2g_{t\phi}\left(\frac{dt}{d\tau}\right)\left(\frac{d\phi}{d\tau}\right)
                    +g_{\phi\phi}\left(\frac{d\phi}{d\tau}\right)^2\nn\\
                    &&+ g_{\rho\rho}\left(\frac{d\rho}{d\tau}\right)^2
                    +g_{zz}\left(\frac{dz}{d\tau}\right)^2 \label{4momentum}\eea
In this equation, one can define the angular velocity $\Omega\equiv\frac{d\phi}{dt}$. Then for the circular and equatorial orbits, eq.~(\ref{4momentum}) defines the redshift factor between coordinate and proper time,
\be \left(\frac{d\tau}{dt}\right)^2=-g_{tt}-2g_{t\phi}\Omega -g_{\phi\phi}\Omega^2, \label{redshift}\ee
and the energy and the angular momentum take the form,
\begin{align} \tilde{E}\equiv \frac{E}{\mu}&=\frac{-g_{tt}-g_{t\phi}\Omega}{\sqrt{-g_{tt}-2g_{t\phi}\Omega -g_{\phi\phi}\Omega^2}},\label{energy}\\
         \tilde{L}\equiv \frac{L}{\mu}&=\frac{g_{t\phi}+g_{\phi\phi}\Omega}{\sqrt{-g_{tt}-2g_{t\phi}\Omega -g_{\phi\phi}\Omega^2}}, \end{align}
where we have introduced the energy and angular momentum per unit mass. 
From the conditions, $\frac{d\rho}{dt}=0,\,\frac{d^2\rho}{dt^2}=0,\,z=0$ and $\frac{dz}{dt}=0$ for circular equatorial orbits, and the equations of motion
obtained assuming the Lagrangian, ${\cal L}=\frac{1}{2}g_{ab}\dot{x}^a\dot{x}^b$, the orbital angular velocity can be calculated to be,
\be
\Omega=\frac{-g_{t\phi,\rho}+\sqrt{(g_{t\phi,\rho})^2-g_{tt,\rho}g_{\phi\phi,\rho}}}{g_{\phi\phi,\rho}}.
\ee
This is the orbital frequency of a particle in a circular orbit on the equatorial plane and the commas indicate partial derivatives with respect to the corresponding coordinates.            

Equation (\ref{4momentum}) can take a more general form in terms of the constants of motion,
\bea -g_{\rho\rho}\left(\frac{d\rho}{d\tau}\right)^2 -g_{zz}\left(\frac{dz}{d\tau}\right)^2\!\!\!\!\!&=&\!\!\!\!\!
    1-\frac{\tilde{E}^2g_{\phi\phi}+2\tilde{E}\tilde{L} g_{t\phi}+\tilde{L}^2g_{tt}}{(g_{t\phi})^2-g_{tt}g_{\phi\phi}}\nn\\
     \!\!\!\!\!&=&\!\!\!\!\!V_{eff}.\label{eqmotion}\eea
With equation (\ref{eqmotion}) we can study perturbations around circular equatorial orbits. If we assume small deviations of the form, $\rho=\rho_c+\delta\rho$ and $z=\delta z$, then we obtain the perturbed form of equation (\ref{eqmotion}),

\bea -g_{\rho\rho}\left(\frac{d(\delta\rho)}{d\tau}\right)^2 -g_{zz}\left(\frac{d(\delta z)}{d\tau}\right)^2\!\!\!\!\!&=&\!\!\!\!\!
       \frac{1}{2}\frac{\p^2 V_{eff}}{\p\rho^2} (\delta \rho)^2 \nn\\
       \!\!\!\!\!&&\!\!\!\!\!+  \frac{1}{2}\frac{\p^2 V_{eff}}{\p z^2} (\delta z)^2. \eea
This equation describes two harmonic oscillators with frequencies,
\begin{align} \bar{\kappa}_{\rho}^2&=\left.\frac{g^{\rho\rho}}{2}\frac{\p^2 V_{eff}}{\p\rho^2}\right|_c\,,\\
                                 \bar{\kappa}_z^2&=\left.\frac{g^{zz}}{2}\frac{\p^2 V_{eff}}{\p z^2}\right|_c\, ,\end{align}
where everything is evaluated at the corresponding equatorial circular orbit. The first frequency is the frequency of radial oscillations around the radius of the circular orbit, while the second frequency is the oscillation over and under the equatorial plane. The differences of these frequencies (corrected with the redshift factor (\ref{redshift}), i.e., $\kappa_a=(d\tau/dt)\bar{\kappa}_a$) from the orbital frequency,
$\Omega_a=\Omega-\kappa_a$, define the precession frequencies, where the oscillation frequencies $\kappa_a$ are given in terms of the metric functions as,
\bea
\kappa_a^2&=&-\frac{g^{aa}}{2}\left\{(g_{tt}+g_{t\phi}\Omega)^2\left(\frac{g_{\phi\phi}}{\rho^2}\right)_{,aa}\right.\nn\\
              &&-2(g_{tt}+g_{t\phi}\Omega)(g_{t\phi}+g_{\phi\phi}\Omega)\left(\frac{g_{t\phi}}{\rho^2}\right)_{,aa}\\
              &&\left.+(g_{t\phi}+ g_{\phi\phi}\Omega)^2\left(\frac{g_{tt}}{\rho^2}\right)_{,aa}\right\}\Bigg |_{z=0}\,,\nn
\eea        
where the index $a$ takes either the value $\rho$ or $z$ to express the frequency of the radial or the vertical perturbation
respectively and the expressions are evaluated on the equatorial plane $z=0$. The position where $\kappa_{\rho}^2$ becomes zero is the location of the ISCO. 

In what follows we are going to look into corotating orbits and all the relevant discussion will be on this type of orbits.

\subsection{Location of the ISCO}

The first property of the geodesics around a NS that we will examine is the location of the ISCO. The ISCO is located at the position where $\kappa_{\rho}^2=0$. Assuming the two-soliton analytic spacetime with the moments chosen as it was described in section \ref{sec:2}, utilizing the universal relations between the moments, we can have by solving $\kappa_{\rho}^2=0$ the location of the ISCO for a general NS spacetime independently of any specific EoS. 

Using as free parameters the spin parameter $j$ and the quadrupolar deformability parameter $\alpha$, we can calculate the reduced circumferential radius of the ISCO, $R_{ISCO}/M$. 
%
\begin{figure}
\centering
\includegraphics[width=.4\textwidth]{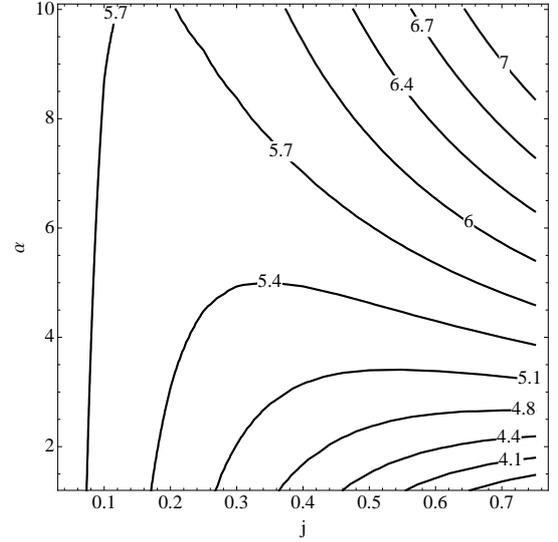}
\caption{The figure shows contours of constant $R_{ISCO}/M$ on the parameter plane of $(j,\alpha)$. The range of the parameters $(j,\alpha)$ has been chosen so as to include all physically relevant NS models, i.e., models with masses in the range between the maximum stable mass and a minimum of around $1M_{\odot}$ with rotation from zero up to the maximum allowed limit. 
}
\protect\label{RiscoPlot}
\end{figure}
%
Fig. \ref{RiscoPlot} shows the contours of constant $R_{ISCO}/M$. Something that we should point out here is that the contour of $R_{ISCO}/M=6$, as the spin parameter $j$ decreases, goes up asymptotically to the axis of $\alpha$ and when $j=0$ it spans the entire axis of $\alpha$. 

\begin{figure}
\centering
\includegraphics[width=.4\textwidth]{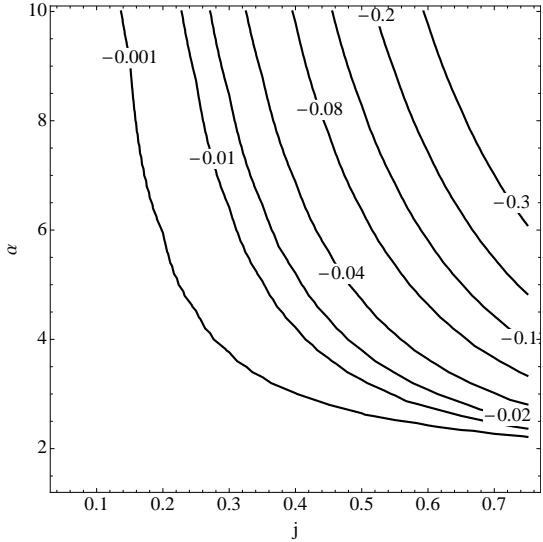}
\caption{The figure shows contours, on the parameter plane of $(j,\alpha)$, of constant relative difference between the $R_{ISCO}$ of the spacetime and the $R_{ISCO}^{ss}$ calculated with the Shibata~\&~Sasaki formula up to third order in the expansion, i.e., up to the order that includes the spin octupole.
}
\protect\label{RiscoSS}
\end{figure}

One thing that would be interesting to explore here, since we have the EoS-independent ISCO, is the generic accuracy of the formula developed by \cite{SSisco} that gives the radius of the ISCO as an expansion on the multipole moments.  The formula developed by Shibata~\&~Sasaki gives the ISCO radius as an expansion in the multipole moments of the spacetime going up to the mass hexadecapole, $M_4$. Since we are using here expressions that are taking consistently into account the moments up to the spin octupole, $J_3$, and since \cite{berti-stergioulas} have demonstrated that it is better to be consistent in the order of the expansion that is used in the Shibata~\&~Sasaki formula, we will use the formula up to the third order in the expansion, i.e.,

\bea  R_{ISCO}^{ss}&=&6M(1 - 0.54433 j - 0.22619 j^2 + 0.17989 \alpha j^2  \nn\\
                 &&- 0.23002 j^3 + 0.26296 \alpha j^3 - 0.05317 j^3 \beta),
\eea
where $\alpha$ is the reduced quadrupole in our formalism and $\beta\equiv-J_3/(j^3M^4)$ is the reduced spin octupole. The parameter $\beta$ in this formula will be given in terms of the parameter $\alpha$, as it was discussed at the end of section \ref{sec:2}, having therefore the ISCO in our two dimensional parameter space, $(j,\alpha)$. In Fig. \ref{RiscoSS} we show a contour plot of the relative difference between the $R_{ISCO}$ and the $R_{ISCO}^{ss}$, $\Delta R_{ISCO}= (R_{ISCO}-R_{ISCO}^{ss})/R_{ISCO}$. 
%
%
As one can see from the plot, the Shibata~\&~Sasaki formula is very accurate for small values of the spin parameter $j$ and it remains accurate for a large part of the parameter space. The relative difference goes over 10 per cent only on the upper right corner of the parameter space, where we have the rapidly rotating NSs that have also a relatively large deformability. This is the behaviour that one would expect, since for these models the deformation is large enough to make the higher order moments more relevant for the calculation of the ISCO.

One last thing that we should consider at this point is for which part of the parameter space, the position of the ISCO is outside of the surface of the star, i.e., for which region of the parameter space the equatorial circumferential radius of the stellar surface $R_{eq}$ is smaller than the $R_{ISCO}$. In the appendix \ref{radii} we discuss how one could have an EoS-independent description of the equatorial radius of a NS in terms of the two dimensional parameter space that we are using. Fig. \ref{surfacecontour} gives the reduced circumferential equatorial radius of a NS in terms of the parameter space $(j,\alpha)$. What we would like to see is for which part of the parameter space the inequality $R_{eq}<R_{ISCO}$ holds. 

\begin{figure}
\centering
\includegraphics[width=.4\textwidth]{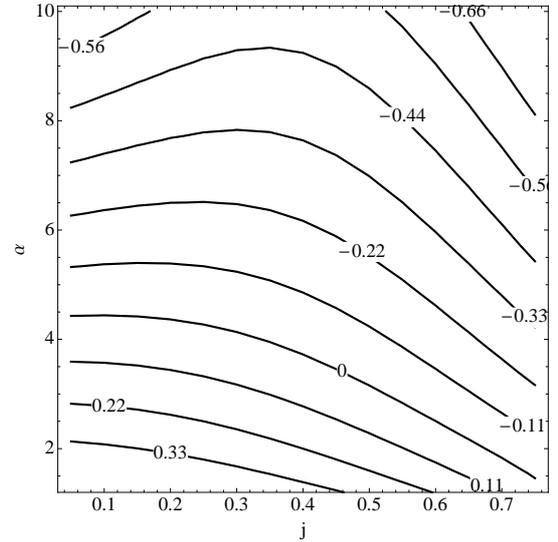}
\caption{The figure shows contours, on the parameter plane of $(j,\alpha)$, of constant relative difference between the $R_{ISCO}$ of the spacetime and the equatorial circumferential radius of the star $R_{eq}$, i.e., $(R_{ISCO}-R_{eq})/R_{ISCO}$. The contours with a positive value correspond to a region of the parameter space where the stars have their surfaces under the ISCO. The negative values correspond to stars that have engulfed their ISCO. 
}
\protect\label{RiscoSurface}
\end{figure}

For that reason we plot in Fig. \ref{RiscoSurface} the relative difference between the $R_{ISCO}$ of the spacetime and the equatorial circumferential radius of the star $R_{eq}$, i.e., $(R_{ISCO}-R_{eq})/R_{ISCO}$. Positive values correspond to models that have their ISCO outside the stellar surface, while negative values correspond to models that have their ISCO buried under the surface of the star. Fig. \ref{RiscoSurface} shows that when the deformability of a star is large enough, then the ISCO is buried under the surface. But as we can see by comparing Figs \ref{RiscoSurface} or \ref{surfacecontour} against Fig. \ref{RiscoPlot}, the radius of the star is not that much larger than the ISCO and it can be up to around 50 per cent of the ISCO, i.e., for a star that has its ISCO in the range of $6-7M$ the surface will be in the range of $10-12M$.  

This is something that one needs to take into account when considering observables from orbiting matter around a compact object. The presence of the surface will constrain the observed orbital and precession frequencies that a fluid element can have as it orbits around a NS and it will also constrain the appearance of an accretion disc that will either terminate at the ISCO or at the surface, depending on the model. 

One final point that one could make here is that, from comparing Figs \ref{RiscoSS} and \ref{RiscoSurface}, one can conclude that for the part of the parameter space that an ISCO exists outside the NS, the Shibata~\&~Sasaki formula is extremely accurate (better than 1 per cent). 

\subsection{Orbital frequency $\Omega_{ISCO}$ at the ISCO}

\begin{figure}
\centering
\includegraphics[width=.4\textwidth]{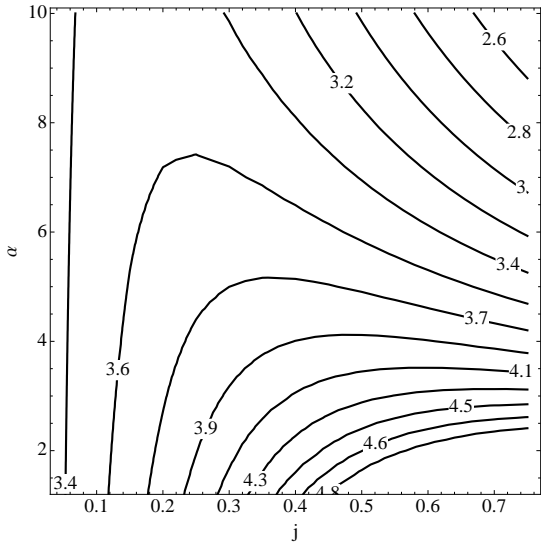}
\caption{The figure shows contours, on the parameter plane of $(j,\alpha)$, of constant reduced frequency $M\times\nu_{ISCO}=M\times\Omega_{ISCO}/(2\pi)$ in units of $\textrm{km}\times\textrm{kHz}$. 
}
\protect\label{viscofig1}
\end{figure}

The second property of the geodesics that can be a potential astrophysical observable is the orbital frequency at the ISCO, $\Omega_{ISCO}$, or rather the corresponding frequency $\nu=\Omega/(2\pi)$. One would expect that this frequency might appear in QPOs associated to the orbital motion of fluid elements in the accretion disc. In particular it might have distinctive characteristics, since at that point the accreting material plunges to the compact object and the disc terminates. In Fig. \ref{viscofig1} we have plotted contours of the normalized frequency $M\times\nu_{ISCO}$ in units of $\textrm{km}\times\textrm{kHz}$. 

\begin{figure}
\centering
\includegraphics[width=.4\textwidth]{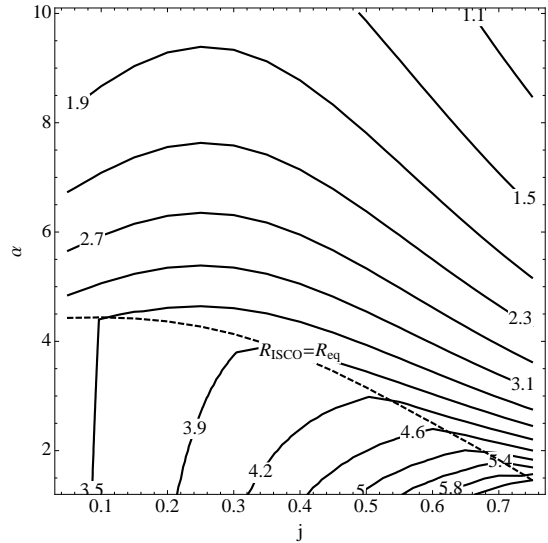}
\caption{The figure shows contours, on the parameter plane of $(j,\alpha)$, of constant reduced frequency $M\times\nu_{K}^{max}$ in units of $\textrm{km}\times\textrm{kHz}$ The dashed line that intersects the solid lines represents the models for which the ISCO radius is equal to the surface radius. Below that curve the orbital frequency plotted is that of the ISCO which is outside the surface of the star. Above that curve the orbital frequency plotted corresponds to the orbital frequency at the surface of the star, since the ISCO is under the surface. 
}
\protect\label{viscofig2}
\end{figure}

In the previous section though, we talked about the location of the ISCO with respect to the surface, and we saw that for some models it is possible that the surface of the compact object overcomes the location of the ISCO. This results to having the available space for orbital motion terminate before the ISCO, at the radius of the surface. This means that since we have to take also into account the surface, the figure for the frequency will be modified. We can therefore plot essentially the maximum orbital frequency, $\nu_K^{max}$, which is the orbital frequency of the closest stable circular orbit to the surface of the compact object. The relevant contour plot is given in Fig. \ref{viscofig2}, where we also indicate the models for which the ISCO radius is equal to the surface radius, $R_{ISCO}=R_{eq}$. Below that curve the orbital frequency plotted is that of the ISCO, while over that curve the orbital frequency plotted corresponds to the orbital frequency at the surface of the star.

It is tempting at this point to create a combined figure of the contours for the maximum orbital frequency and the rotation frequency of a NS, presented in the appendix \ref{NSrotation}. Both of these frequencies are potentially observable and are probably the most prominent observables one might have from an accreting rotating NS. We present this figure at this point because it displays an interesting physical characteristic of rotating NSs which this parameterization nicely demonstrates. 

\begin{figure}
\centering
\includegraphics[width=.4\textwidth]{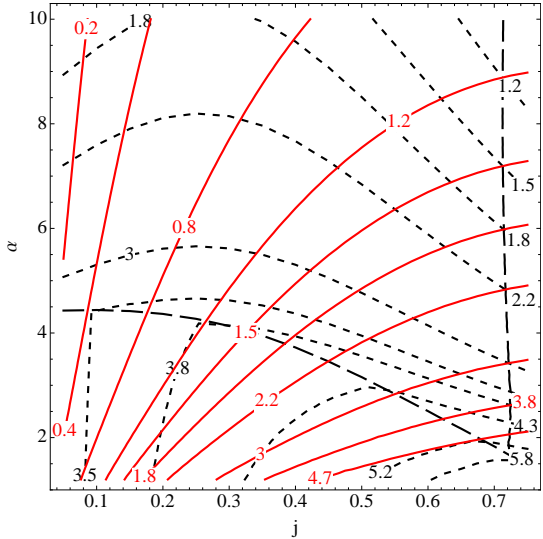}
\caption{The figure shows the contours, on the parameter plane of $(j,\alpha)$, of constant reduced frequency $M\times\nu_{K}^{max}$ (dotted curves) overlaid against the contours of the rotation frequency of NSs (solid curves), in units of $\textrm{km}\times\textrm{kHz}$. The figure shows also the dashed curve of $R_{ISCO}=R_{eq}$ and another vertical dashed curve that represents the models that are rotating at the Kepler limit. This becomes evident if one notices, that the curves of constant stellar rotational frequency intersect the curves of constant orbital frequency at the surface, that have the same value. The region to the right of that curve corresponds to physically not realisable stellar models.
}
\protect\label{viscofig3}
\end{figure}

Fig. \ref{viscofig3} shows the contours of constant reduced maximum frequency $M\times\nu_{K}^{max}$ overlaid against the contours of constant rotation frequency of NSs. The interesting feature of this figure is the vertical dashed curve that is located in the region above the curve of $R_{ISCO}=R_{eq}$, i.e., in the region where the dotted curves of maximum orbital frequency correspond to the orbital frequency at the radius of the surface. One notices that along the vertical dashed curve, curves of constant NS rotation frequency intersect curves of the same constant orbital frequency. This signifies the models that are rotating at the Kepler limit, i.e., at a frequency such that a fluid element on the surface of that star would have to rotate at a Keplerian orbital frequency. This means that the fluid elements at the surface of such a configuration are not gravitationally bound to the star and all models to the right of the vertical dashed curve, i.e., models that rotate more rapidly than the Keplerian orbital frequency, can not be physically realized. The vertical dashed curve therefore represents a physical boundary to our parameter space for NSs, the boundary of maximally rotating models. It is interesting to note that the Kepler limit is at $j\sim0.72$ and is quite independent of the parameter $\alpha$. We should also note that since our description applies for all EoSs, the limit that we get for the spin parameter is EoS-universal. This result is reminiscent of the universal empirical formula for the Keplerian frequency by \cite{2004Sci...304..536L,2009A&A...502..605H}, but it is not clear how one could get from that formula to this result.

\subsection{Vertical oscillation frequency $\kappa_{ISCO}$ at the ISCO}

\begin{figure}
\centering
\includegraphics[width=.4\textwidth]{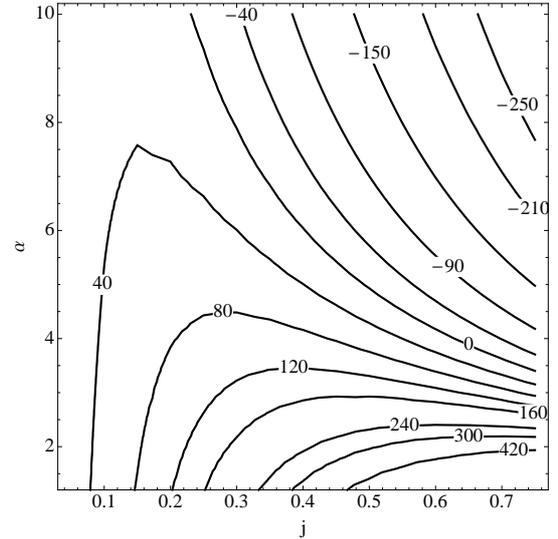}
\caption{The figure shows contours, on the parameter plane of $(j,\alpha)$, of constant reduced nodal precession frequency $M\times(\Omega_z/2\pi)_{ISCO}$ in units of $\textrm{km}\times\textrm{Hz}$. 
}
\protect\label{OmegaZisco}
\end{figure}

Another geodesic property that we could associate with the location of the ISCO is the nodal precession of a slightly off-equatorial orbit, $\nu_z=(\Omega-\kappa_z)/(2\pi)$ at the location of the ISCO. The behaviour of the nodal precession frequency of NSs is quite different from the behaviour of the corresponding Kerr BHs. In particular the nodal precession exhibits some very interesting features that can not be found in rotating BHs and essentially demonstrates the fact that the exterior spacetime of NSs deviates from that of Kerr both quantitatively and qualitatively. To demonstrate this, we are plotting in Fig. \ref{OmegaZisco} the normalized nodal precession frequency, $M\times\nu_z$ at the location of the ISCO. 

The feature that stands out is the fact that the nodal precession is positive at the ISCO for some models while for other models it is negative. This effect, which had been noticed by \cite{PappasQPOs} in connection to QPO observations and NS properties, is further discussed in appendix \ref{app:Nodal}. This means that if there were a slightly inclined accretion disc around a NS that demonstrated negative precession frequency at the ISCO, then that accretion disc would have the inner region precessing in one direction while the outer region would precess in the opposite direction (see discussion in appendix \ref{app:Nodal}). This effect could be of significance to the modelling of accretion discs and could have even more spectacular consequences than the effects explored by \cite{King2012ApJ} for accreting BHs. 

\begin{figure}
\centering
\includegraphics[width=.4\textwidth]{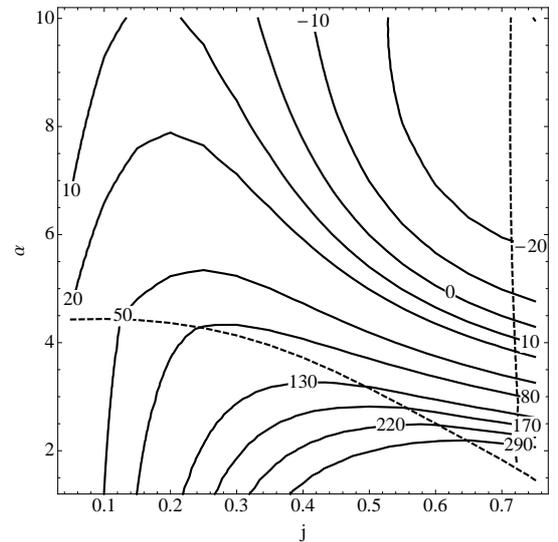}
\caption{Same plot as in Fig. \ref{viscofig2}, but for the nodal precession frequency. Again the horizontal dashed line represents the models for which $R_{ISCO}=R_{eq}$, while the vertical dashed line represents the models at the Kepler limit. 
}
\protect\label{OmegaZisco2}
\end{figure}

In this case as well as in the previous cases, one needs to take into account the effect that the surface of the NS will have, since the presence of the surface in some cases will not allow matter to access radii all the way down to the ISCO. Therefore, if we focus only on accessible precession frequencies the previous picture will be modified. In Fig. \ref{OmegaZisco2} we show the contours of constant nodal precession frequency with the modification that above the horizontal dashed line of $R_{ISCO}=R_{eq}$, the frequency corresponds to the location of the surface. We can see from Fig. \ref{OmegaZisco2} that even though the presence of the surface changes things, there are still models that can have a negative precession frequency at the inner part of the spacetime close to the surface of the star.   

\subsubsection{Region of zero nodal precession}

\begin{figure}
\centering
\includegraphics[width=.4\textwidth]{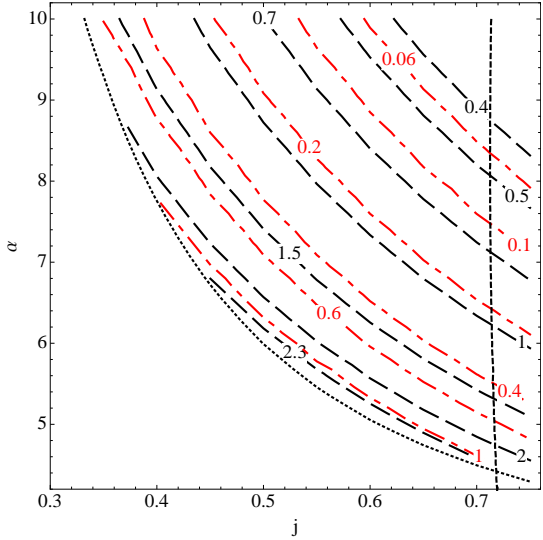}
\caption{Same plot as in Fig. \ref{viscofig2}, but for the orbital frequency (black long dashed curve) and the periastron precession frequency (red dash-dotted curve) at the location where the nodal precession becomes zero. The frequencies are in units of $\textrm{km}\times\textrm{kHz}$. The almost diagonal dotted line indicates the models for which the radius of the surface coincides with the radius where the nodal precession is zero, while the vertical dashed line represents the models at the Kepler limit. Also, the vertical axis has been modified in order to include only the relevant region of the figure.
}
\protect\label{OmegaZisco3}
\end{figure}

The effect discussed in the previous section, i.e., that the nodal precession near the surface of the star is negative, means that there is a radius further out from the surface where the nodal precession is zero. In the case that we have an accretion disc, that region is a place where there might be something significant taking place, such as the breaking of the disc for example. Therefore it would be of interest to see what is the behaviour at that point of the other two frequencies, i.e., the orbital frequency and the periastron precession frequency. Contours of these two frequencies are plotted in Fig. \ref{OmegaZisco3}. The dotted diagonal line indicates the models for which the radius of the surface coincides with the radius where the nodal precession is zero. Below that curve there is no zero nodal precession outside the star.

\subsection{Thin accretion disc}
\label{sec:thinDisc}

The relativistic model of a radiatively efficient thin accretion disc was developed by \cite{Novikov-Thorne1973} and \cite{Page-Thorne1974}. The disc is thin in the sense that the vertical size of the disc, $H$ is much smaller than the characteristic radial length of the disc. It is also radiatively efficient in the sense that all the gravitational energy lost by the accreting matter due to viscous stresses is efficiently radiated away or transported outwards, therefore there is no heating of the disc. The disc is also considered to be in a hydrodynamic equilibrium and in a steady state where the accretion rate $\dot{M}_0$ is constant. Furthermore, the pressure gradient and the heat flow within the disc are negligible in the radial direction, while heat can flow in the vertical direction (for details on the assumptions see \cite{Page-Thorne1974}). 

This construction results in an accretion disc where the plasma moves in equatorial Keplerian orbits with angular velocity $\Omega$, energy per unit mass $\tilde{E}$ and angular momentum per unit mass $\tilde{L}$, slowly inspiraling towards the central object releasing gravitational energy that is radiated from the two faces of the disc. In the end, the structure equations of the disc are given by the mass conservation, the energy conservation and the angular momentum conservation. In this construction, everything is integrated along the thickness of the disc and the metric is assumed to be on the equatorial plane $z=0$, while deviations from the equatorial plane behave as $(z/\rho)^2$, due to equatorial symmetry, and are considered negligible.

The energy momentum tensor for the accreting matter fluid has the form, 
\be T^{\mu\nu}=\rho_0u^{\mu}u^{\nu}+2u^{(\mu}q^{\nu)}+t^{\mu\nu} \ee
where $u^{\mu}$ is the four-velocity of particles, $q^{\mu}$ is the energy flow four-vector, $\rho_0$ is the rest mass density, and $t^{\mu\nu}$ is the stress tensor (the specific internal energy $\Pi$ is negligible according to the initial assumptions). The first structure equation comes from the conservation of the rest mass, $\nabla_{\mu}(\rho_0u^{\mu})=0$, and it essentially states the fact that the time-averaged mass accretion rate is independent of the radius, i.e.,
\be \dot{M}_0\equiv-2\pi\sqrt{-g}\Sigma u^{\rho}=\textrm{const.}, \ee 
where $g$ is the determinant of the metric (in our case the metric \eqref{Pap}, which is calculated on the equatorial plane), $u^{\rho}$ is the radial component of the four-velocity, and $\Sigma$ is the surface density defined as, $\Sigma\equiv\int_{-H}^{H} \langle\rho_0 \rangle dz$, where $ \langle \rho_0 \rangle$ is the rest mass density averaged in time and over the azimuthal angle $\phi$.

The second equation of structure comes from the angular momentum conservation and is,
\be \left[\dot{M}_0\tilde{L}-2\pi\sqrt{-g}W_{\phi}^{\rho}\right]_{,\rho}=4\pi\sqrt{-g}F\tilde{L}, \ee
where $F$ is the radiated energy flux from one of the surfaces of the disc and $W_{\phi}^{\rho}$ is the time-averaged torque per unit circumference at a particular radius, due to the stresses. The latter is defined as $W_{\phi}^{\rho}(\rho)\equiv\int_{-H}^{H} \langle t_{\phi}^{\rho} \rangle dz$. The last equation of structure comes from the energy conservation and is,
\be \left[\dot{M}_0\tilde{E}-2\pi\sqrt{-g}\Omega W_{\phi}^{\rho}\right]_{,\rho}=4\pi\sqrt{-g}F\tilde{E}. \ee
To these equations we need to add the fundamental relation for circular geodesics, i.e., 
\be \tilde{E}_{,\rho}=\Omega\tilde{L}_{,\rho}.\ee
By making the redefinitions 
\be f \equiv 4\pi\sqrt{-g}F/\dot{M}_0,\quad w \equiv 2\pi\sqrt{-g}W_{\phi}^{\rho}/\dot{M}_0, \ee
the angular momentum and energy conservation equations can take the form 
\be (\tilde{L}-w)_{,\rho}=f \tilde{L}, \quad  (\tilde{E}-\Omega w)_{,\rho}=f \tilde{E}. \ee
From these two equations one can obtain the algebraic relation for $w$, 
\be w=\frac{\tilde{E}-\Omega \tilde{L}}{(-\Omega_{,\rho})} f, \ee
and by substituting this, in the previous equations, we reach after some algebra to the integrable equation for $f$,
\be \left(\frac{(\tilde{E}-\Omega \tilde{L})^2}{(-\Omega_{,\rho})} f \right)_{,\rho} = (\tilde{E}-\Omega \tilde{L}) \tilde{L}_{,\rho}.
\ee
For this equation to be integrated, one needs to choose the appropriate boundary conditions. The usual assumption is that the disc terminates at the ISCO, where matter plunges towards the central object. This means that at the ISCO, which is the inner edge of the disc, one assumes a zero torque condition. With this boundary condition the resulting equation for the flux is 
\be f=-\frac{\Omega_{,\rho}}{(\tilde{E}-\Omega \tilde{L})^2}\int_{\rho_{ISCO}}^{\rho} (\tilde{E}-\Omega \tilde{L}) \tilde{L}_{,\rho'} d\rho' .\ee
As we have seen from our discussion so far, for NSs this is not the only possibility. For some models the surface of the star overcomes the location of the ISCO, therefore in these cases the disc should terminate on the NS surface. Apart from choosing the inner edge of the disc to be at the radius of the surface, we also need to choose and appropriate boundary condition for the torque. As it is discussed  by \cite{Zimmerman2005}, in the region where the accretion disc touches the surface, a boundary layer forms that causes the orbital frequency of the accreting material to decrease until it reaches that of the surface of the rotating star. In that region, we have a maximum of the orbital frequency where there are no viscous shear stresses and therefore there is zero torque. The orbital frequency at the maximum is essentially the maximum geodesic orbital frequency, i.e., the orbital frequency at the radius of the surface. In this case therefore, the result for the flux is 
\be f=-\frac{\Omega_{,\rho}}{(\tilde{E}-\Omega \tilde{L})^2} \int_{\rho_s}^{\rho} (\tilde{E}-\Omega \tilde{L}) \tilde{L}_{,\rho'} d\rho' ,\ee
where the subscript $s$ indicates the values at the surface of the star.

Finally, the emitted flux from the surface of the accretion disc as a function of the radius will be given by the expression 
\be F(\rho)=\frac{\dot{M}_0}{4\pi \sqrt{-g}} f. \ee 
Since the disc is considered to be in thermal equilibrium, the surface of the disc at each radius radiates as a blackbody. As a consequence, the emitted flux is related to the blackbody temperature by the Stefan-Boltzmann law 
\be\label{Stefan-Boltzman} F(\rho)=\sigma T^4(\rho), \ee
where $\sigma$ is the Stefan-Boltzmann constant.\footnote{In the relevant units for us, the constant is $\sigma=1.562\times 10^{-54} \textrm{km}^{-2} \textrm{K}^{-4}$} Therefore from the distribution of the radiated flux we obtain the temperature distribution of the disc.

We should note here how the different quantities scale with the mass of the central object, since we will need it for our analysis. As we have already seen, the radial distances scale proportionally to the mass, $\rho=M\bar{\rho}$, while the frequencies scale as $\Omega=M^{-1} \bar{\Omega}$. Similarly, the orbital energy and angular momentum per unit mass scale as $\tilde{E}=M^0\bar{\tilde{E}}$ and $\tilde{L}=M \bar{\tilde{L}}$. The square root of the determinant of the metric, which is essentially a radius (the metric (\refeq{Pap}) is given in cylindrical coordinates), also scales linearly with the mass. Therefore, the redefined flux $f$ will scale as $f(\rho)=M^{-1}\bar{f}(\bar{\rho})$, while the flux over the mass accretion rate will scale as $F(\rho)/\dot{M}_0=M^{-2} \overline{\left(F(\bar{\rho})/\dot{M}_0\right)}$. From these scalings we have that, modulo the accretion rate, the temperature scales as $T(\rho)=M^{-1/2} \bar{T}(\bar{\rho})$. Actually it would be convenient to add the accretion rate to the scaling, in the sense that we could express the Stefan-Boltzmann law as $F/\dot{M}_0=\sigma \tilde{T}^4$, where $\tilde{T}=(\dot{M}_0)^{-1/4}T$ is an accretion rate reduced temperature, and get the scaling, $T=M^{-1/2}(\dot{M}_0)^{1/4}\bar{T}$. 

\subsection{Emitted efficiency $\eta$ of accretion}
\label{sec:efficiency}

A first measure of the luminosity of the disc can be given by the efficiency $\eta$, defined as

\be \eta=1-\tilde{E}\Big|_{\textrm{in}}, \ee
where $\tilde{E}\Big|_{\textrm{in}}$ is the energy per unit mass of the circular orbit at the inner edge of the accretion disc, which can be either the location of the ISCO or the surface of the star, as we have seen so far. The efficiency $\eta$ is essentially a measure of the energy that has been radiated away to infinity from the disc surface per unit mass of infalling material, assuming that all the photons reach infinity. Since $\tilde{E}$ scales independently of the mass of the central object, the efficiency will also be independent of the mass. 

\begin{figure}
\centering
\includegraphics[width=.4\textwidth]{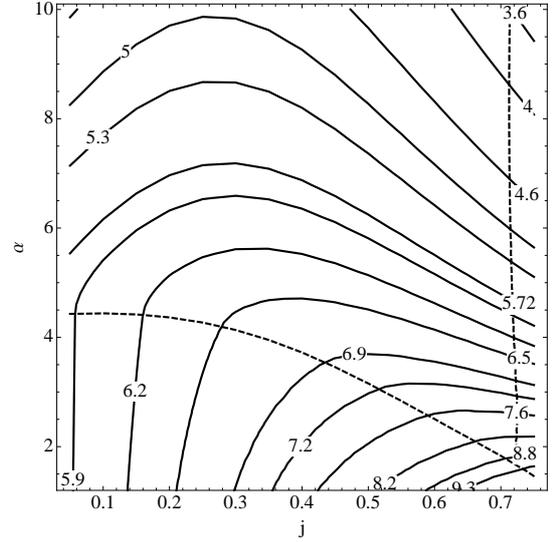}
\caption{Same plot as in Fig. \ref{viscofig2}, but for the efficiency $\eta$ in percentile. The efficiency is independent of the mass of the central object and it is evaluated either at the radius of the ISCO or at the radius of the surface of the star. The almost horizontal dashed line indicates the models for which the radius of the surface coincides with the radius of the ISCO, while the vertical dashed line represents the models at the Kepler limit. 
}
\protect\label{efficiency}
\end{figure}

In Fig. \ref{efficiency} we present the contours of constant efficiency $\eta$, evaluated at the inner edge of the disc. The figure is separated in two parts, the lower part under the dashed curve of $R_{ISCO}=R_{eq}$, where the inner edge of the disc is at the ISCO and the upper part where the inner edge of the disc is at the surface of the star. The values are given per cent. 
One can get the luminosity that corresponds to the given efficiencies by multiplying the efficiency to an accretion rate $\dot{M}_0$.

\subsection{Maximum disc temperature}
%
\begin{figure*}
\centering
\includegraphics[width=.35\textwidth]{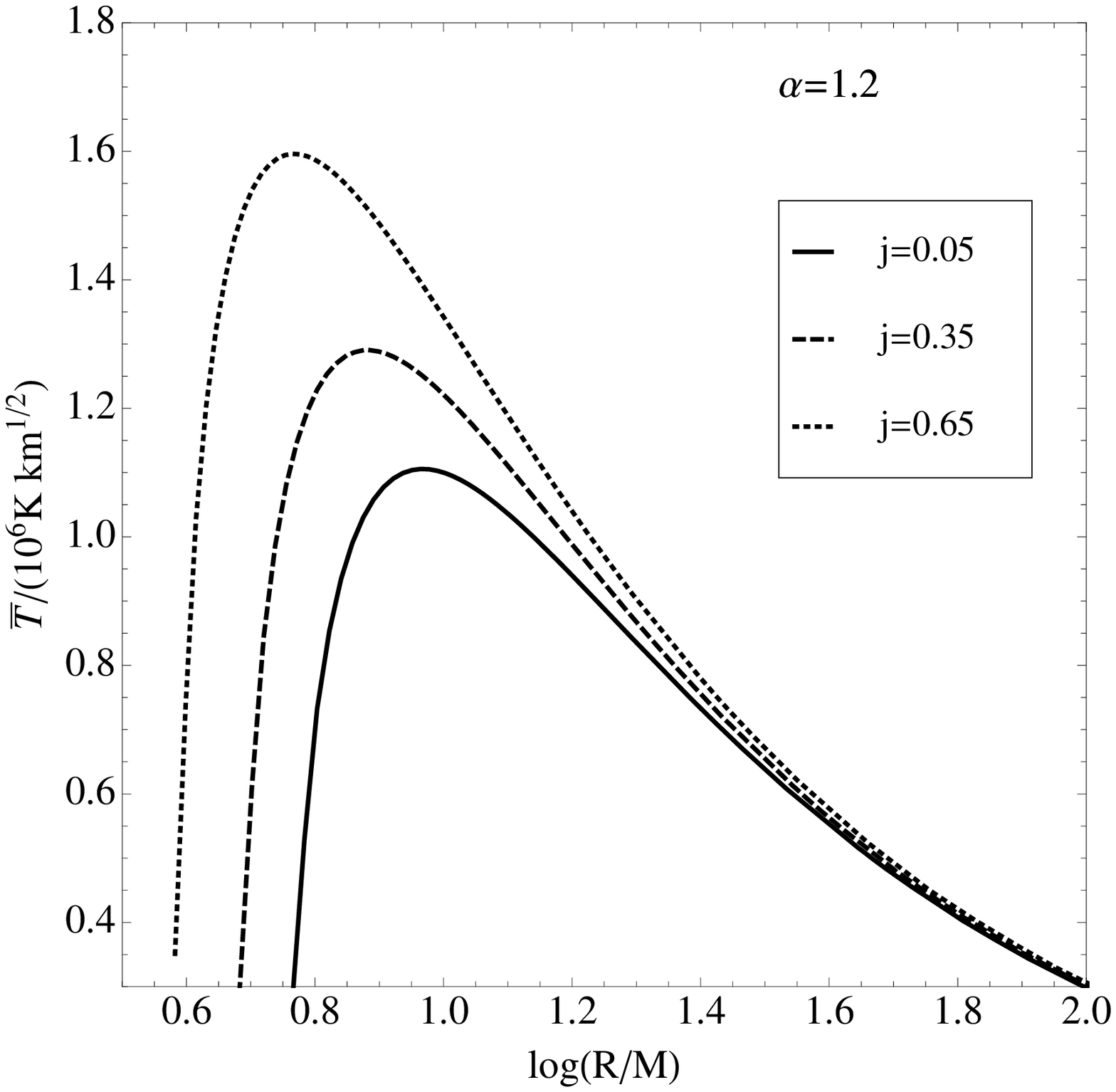} 
\includegraphics[width=.35\textwidth]{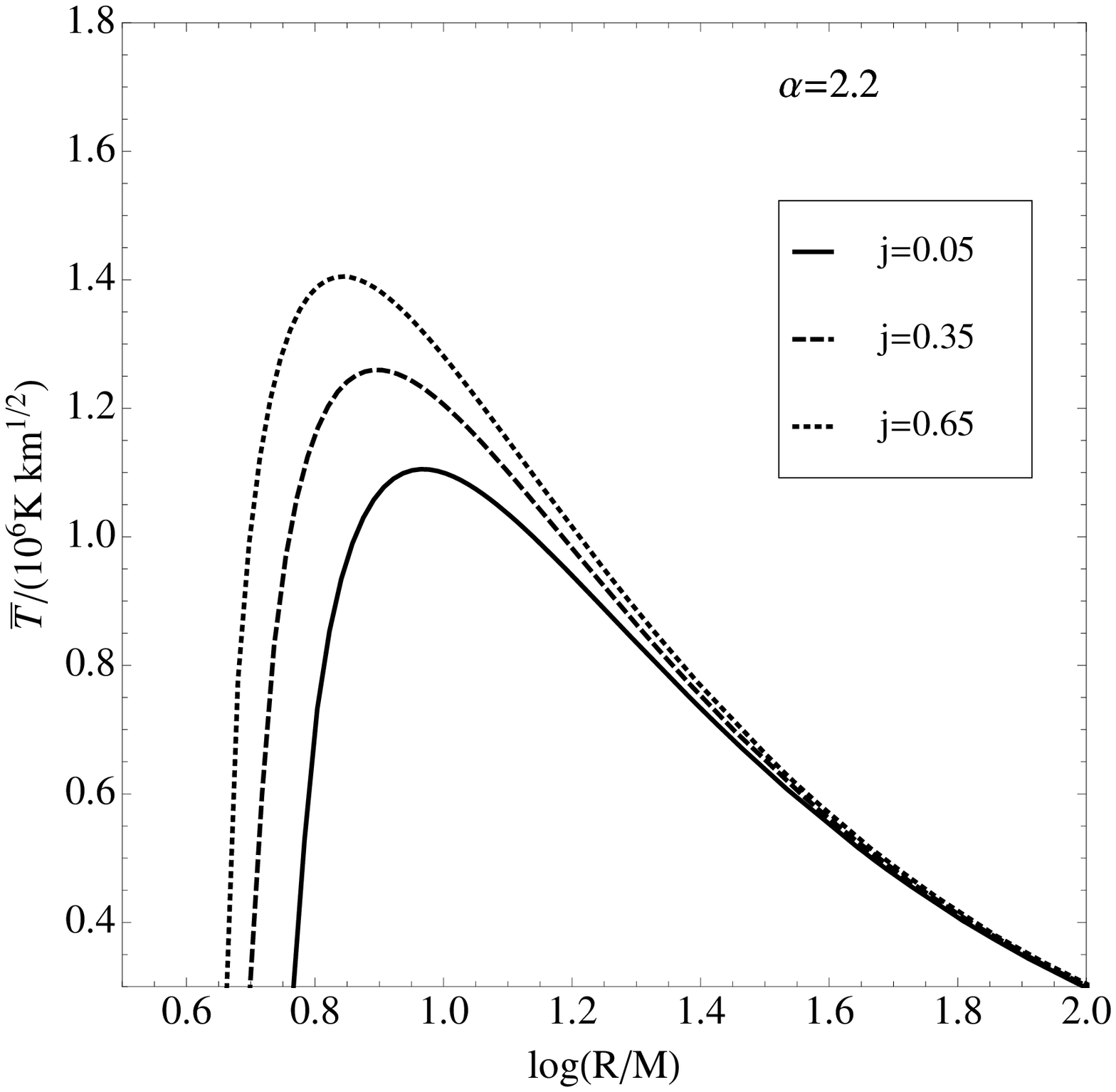}
\includegraphics[width=.35\textwidth]{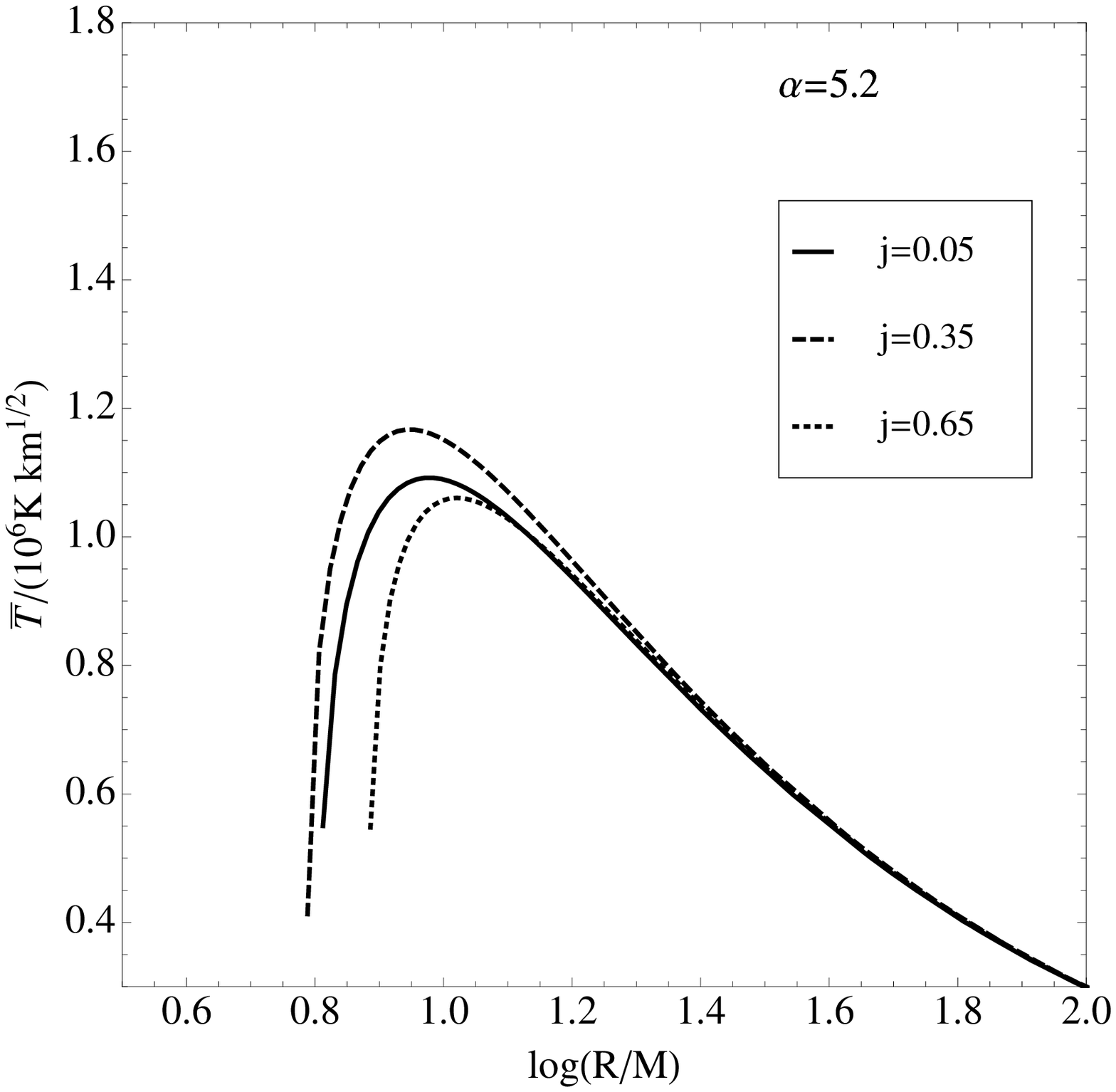} 
\includegraphics[width=.35\textwidth]{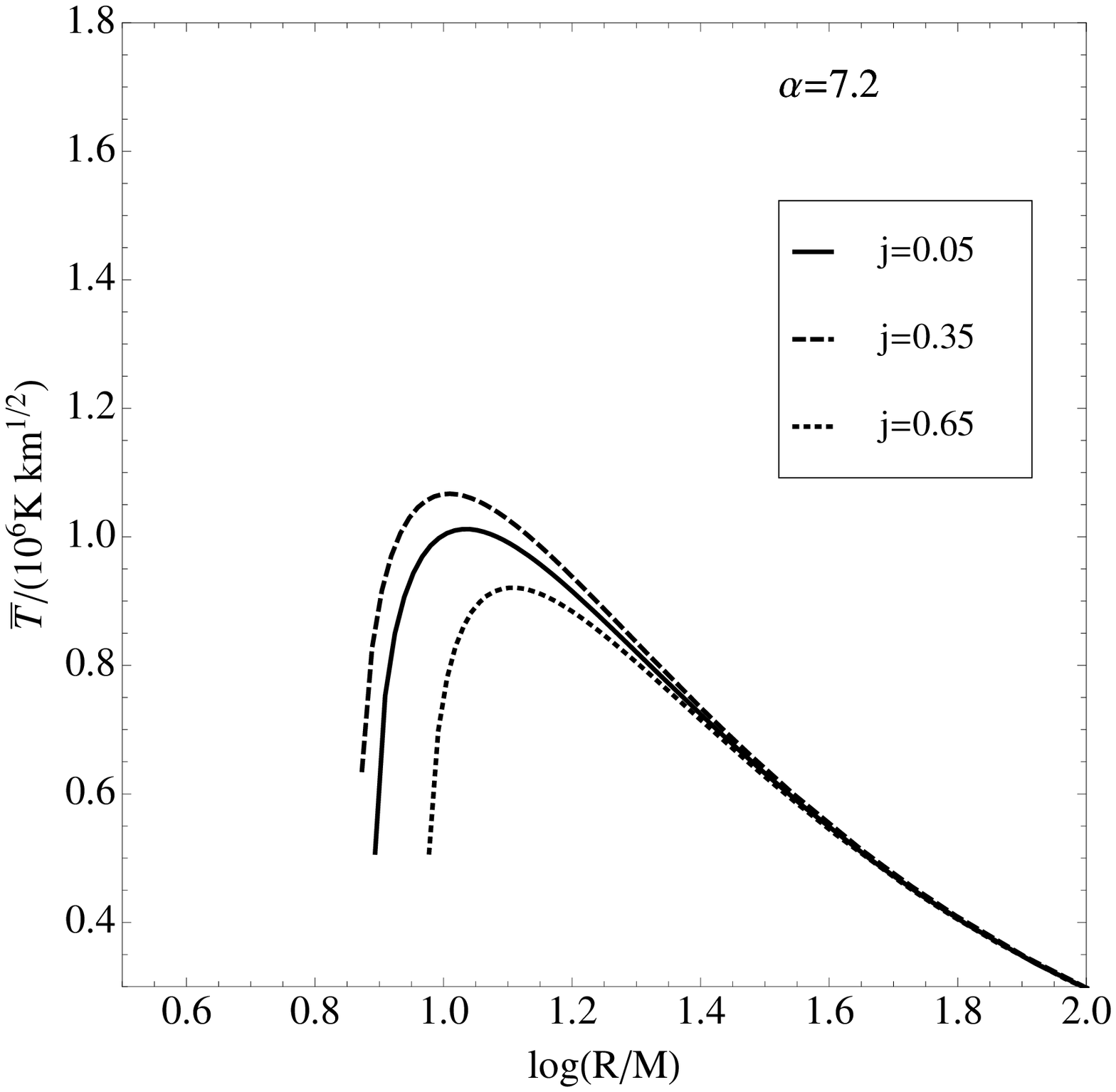}
\caption{Indicative temperature distributions of the accretion discs for different values of the quadrupolar deformation $\alpha$ and the spin parameter $j$. The quantity plotted is the mass-reduced temperature $\bar{T}=M^{1/2}T$ as a function of the logarithm of the reduced circumferential radius $R/M$. One can see that for the values of $\alpha$ and $j$ that correspond to models where the ISCO location is outside the stelar surface (two top panels), the temperature distributions are ordered with rotation. That is because the shape of the distribution and the maximum effective temperature are mainly determined by the location of the inner edge of the disc, which is given by the location of the ISCO, which in turn depends on the rotation. On the other hand, for the models that have values of $\alpha$ and $j$ that correspond to models where the ISCO location is inside the star (two bottom panels), the spin of the NS is not the main factor that determines the shape of the distribution. The main factor in these cases is the location of the surface, which determines the location of the inner edge of the disc. These temperature profiles have been constructed under the assumption of an accretion rate of $\dot{M}_0=10^{-12}M_{\odot} \textrm{year}^{-1}$ as discussed in the main text.  
}
\protect\label{Tprofiles}
\end{figure*}

In the context of the thin disc model, one can calculate the emitted flux profile and from that the temperature distribution for a given model, as we have previously discussed. A quantity that would be characteristic of the emitted spectrum can be considered to be the effective temperature that corresponds to the maximum flux. Therefore we will present here the behaviour of the maximum effective temperature of thin accretion discs around NSs.

We could present the results for the fluxes and the effective temperature distributions in terms of the mass and accretion rate reduced quantities that we discussed in the end of section \ref{sec:thinDisc}, but instead we will prefer to do the calculations with respect to a fiducial accretion rate, which we will chose to be safely below the Eddington limit and close to a plausible outflow value from a companion star in a LMXB. Specifically we will choose $\dot{M}_0=10^{-12}M_{\odot} \textrm{year}^{-1}$ (see \cite{ShakuraSunyaev1973A&A}). The results in this way will be more intuitive. Apart from that we will maintain the mass-reduced quantities. Furthermore, for reasons of more compact and clearer ilustration, in some cases instead of giving the temperature in terms of degrees Kelvin, we will prefer to give it in terms of electron volts by assuming the correspondence of $1\textrm{eV}/k_B=11,604\textrm{K}$ ($k_B$ is the Boltzmann constant). The resulting temperature will be in eV km$^{1/2}$, since the temperatures that we are calculating are mass-reduced. In any case, one can express the temperatures presented here, in terms of any accretion rate and mass of the central object using the following relation
\bea 
\left(\frac{T}{\textrm{eV}}\right)&=& \left(1.477 \frac{M}{M_{\odot}}\right)^{-1/2}\left(\frac{\dot{M}_0}{10^{-12}M_{\odot} \textrm{year}^{-1}}\right)^{1/4}\nn\\
&&\times\left(\frac{\bar{T}}{1\textrm{eV km}^{1/2}}\right).
\eea
This expression can be interpreted to mean that the temperatures presented here are calculated having as a unit of accretion rate the rate of $10^{-12}M_{\odot} \textrm{year}^{-1}$ and as a unit for the mass of the central object, either the geometric unit of 1km or the unit of $(1.477)^{-1}M_{\odot}$.

As it was shown in section \ref{sec:thinDisc}, for a given choice of the spin parameter $j$ and the quadrupolar deformability parameter $\alpha$, which specify a particular geometry for the spacetime, one can calculate the emitted flux of a thin and radiatively efficient disc around a NS as a function of the radius. Then from the Stefan-Boltzmann law \eqref{Stefan-Boltzman} one can obtain the temperature profile of the disc as a function of the radius. Fig. \ref{Tprofiles} shows such temperature profiles for different choices of $\alpha$ and $j$. In particular, assuming the aforementioned accretion rate, the plots show the mass-reduced effective temperature $\bar{T}=M^{1/2}T$ as a function of the reduced circumferential radius $R/M$. The two top panels in Fig. \ref{Tprofiles} correspond to models that have an ISCO, while the two bottom panels correspond to models without an ISCO. What is apparent from comparing the top and the bottom panels is that the shape of the temperature distribution is affected by the location of the inner edge of the disc. In the two upper panels the inner edge is determined by the location of the ISCO and therefore it depends in a specific way on the rotation (changes monotonically with the spin), while in the two bottom panels the inner edge is determined by the location of the surface of the star which changes the behaviour with respect to rotation. The observed behaviour in the latter case becomes clear if one observes Fig. \ref{surfacecontour}, where one can see that for a given value of $\alpha$ and for increasing spin parameter $j$, at first the radius of the star decreases and after a point it starts to increase again. 

\begin{figure}
\centering
\includegraphics[width=.4\textwidth]{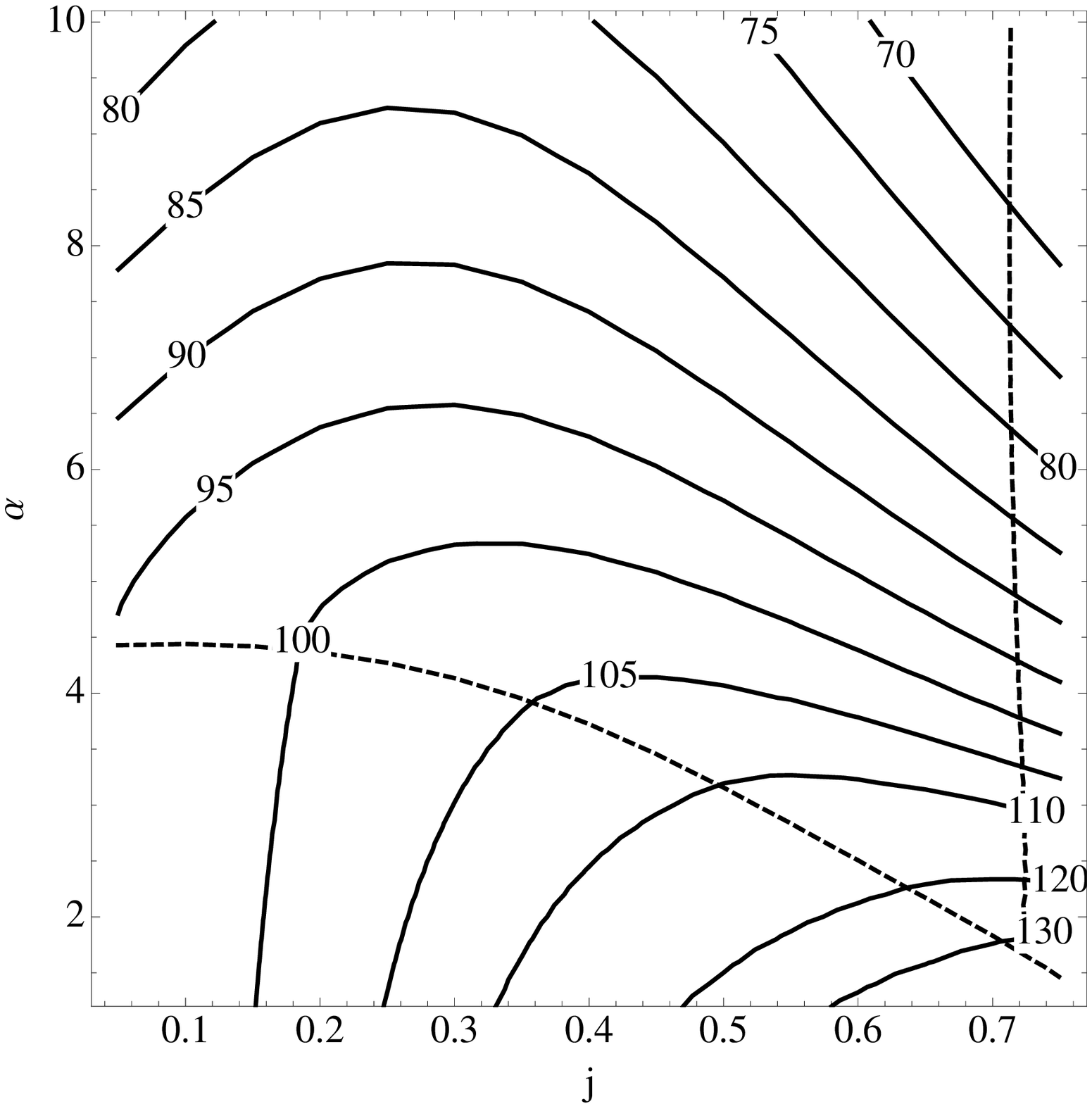}
\caption{Same plot as in Fig. \ref{viscofig2}, but for the maximum mass-reduced effective temperature $\bar{T}_{\textrm{eff}}^{\textrm{max}}$ in eV km$^{1/2}$, under the assumption of an accretion rate of $\dot{M}_0=10^{-12}M_{\odot} \textrm{year}^{-1}$ as discussed in the main text.  
}
\protect\label{maxTeff}
\end{figure}

A characteristic property of these distributions that could be of interest is the maximum effective temperature $\bar{T}_{\textrm{eff}}^{\textrm{max}}$. One could construct contour plots of that temperature in the parameter space of $(j,\alpha)$. %
%
The contours for the maximum effective temperature are given in Fig. \ref{maxTeff}. The plot in Fig. \ref{maxTeff} also shows the curve that separates the models that have the ISCO outside the surface of the star from those that the surface has overcome the ISCO, as well as the limit of maximum rotation of the NSs, as previously discussed.

\begin{figure}
\centering
\includegraphics[width=.4\textwidth]{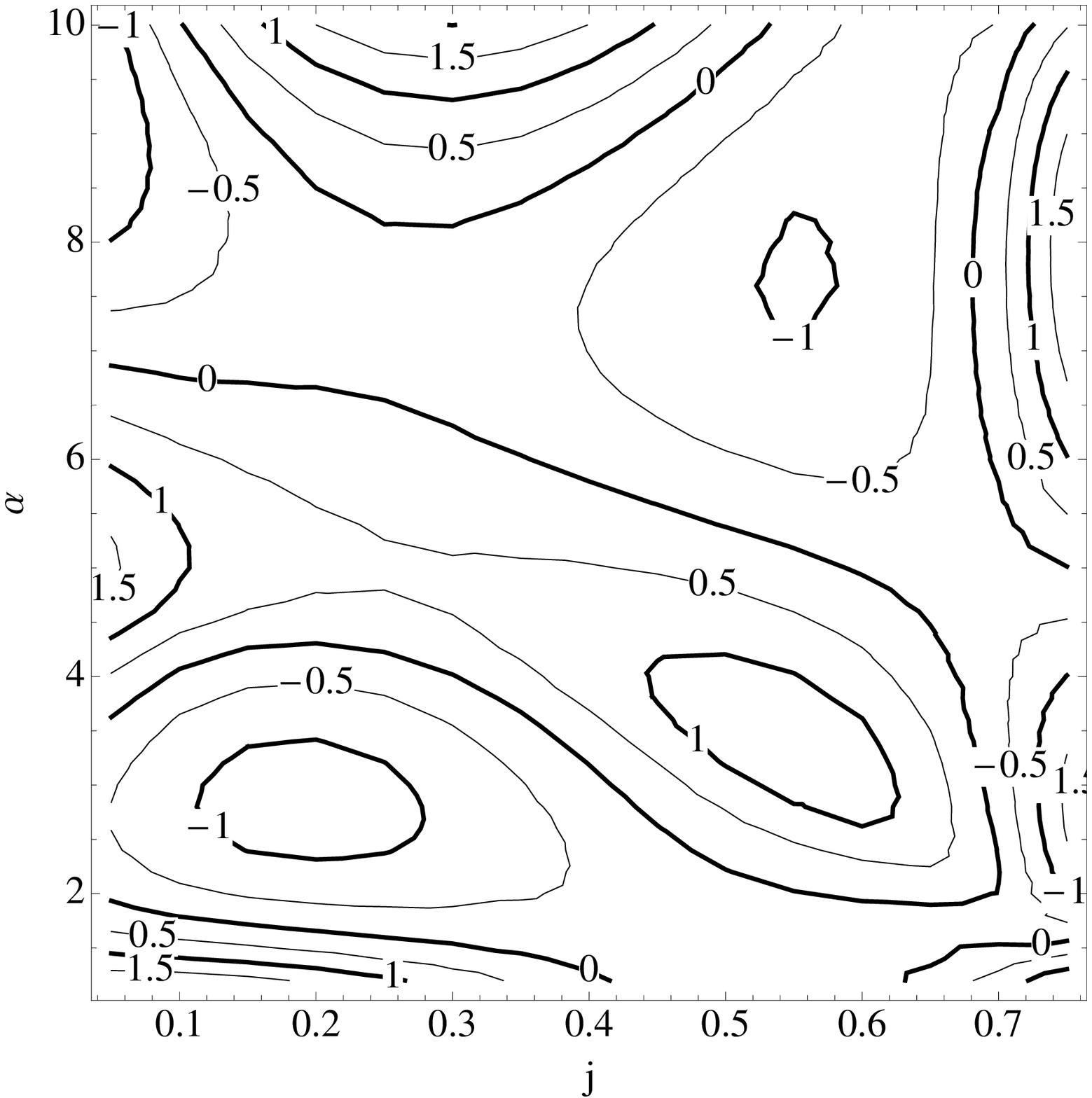}
\caption{Contours of relative difference per cent of the maximum effective temperature with respect to the value given by the fitting function. The absolute value of the relative difference is always smaller than 2 per cent. 
}
\protect\label{TfitError}
\end{figure}

The surface that gives the maximum effective temperature as a function of the spin parameter $j$ and the quadrupolar deformability parameter $\alpha$, can be quite accurately described by a function of the form, 

\bea \bar{T}_{\textrm{eff}}^{\textrm{max}}\!\!\! &=& \!\!\!B_0+B_1 j+B_2 j^2+(A_0+A_1 j+A_2 j^2) \alpha^{n_1}\nn\\
\!\!\!&&\!\!\!+(C_0+C_1 j+C_2 j^2) \alpha^{n_2}, \eea
where $A_0 = -3.95445$, $A_1 = -12.97699$,  $A_2 = 35.90089$, $B_0 = 76.77519$, $B_1 = -44.52375$, $B_2 = 358.23145$,  $C_0 = 18.142198$, $C_1 = 89.3791$,  $C_2 = -322.8577$, $n_1 = 1.107$, and  $n_2 = 0.4425$. In particular, the relative difference between the fitting function and the actual values is always lower than 2 per cent as one can see in Fig. \ref{TfitError}, where we have plotted the relative difference in percentile.


\subsection{Emitted spectrum}

For a given temperature distribution of the accretion disc and the corresponding geometry of the spacetime around the central object, one could calculate the spectrum that a distant observer would measure. A detailed calculation would have to involve various effects, such as that of the propagation of photons in the background spacetime, i.e., light bending, and other effects. One can find a detailed analysis for example in the work by \cite{LietalSpectra2005ApJS}. Here, for illustrative purposes, we will follow a less exhaustive calculation, where we will assume that the disc is almost face on (ignoring inclination effects and light-bending) and we will also neglect non-thermal effects and limb-darkening (neglect colour correction and assume isotropic emission). 
Following this approximation, one can straightforwardly calculate the emitted spectrum of the accretion disc. 

As we have discussed in section \ref{sec:thinDisc}, the accretion disc has a temperature distribution that is given as a function of the radius and the material is assumed to be in local thermodynamic equilibrium and thus the emitted radiation from each location on the surface of the disc is a blackbody radiation. That means that the local specific intensity of the emitted radiation will be
\be I_e(\nu_e)=\frac{2 h \nu_e^3}{c^2} \left[ \exp\left(\frac{h\nu_e}{k_B T_{\textrm{eff}}}\right)-1 \right]^{-1}, \ee  
where $\nu_e$ is the emitted photon frequency and $h$ is the Planck constant. For an observer that is very far from the disc, the observed specific flux will be  
\be F_{\nu_{obs}}=\int I_{obs}(\nu_{obs}) d\Omega_{obs}, \ee
where $I_{obs}$ is the observed specific intensity, $\nu_{obs}$ is the observed photon frequency and $d\Omega_{obs}$ is the observed differential solid angle. The quantities at the observer's frame can be calculated in terms of the quantities on the emission frame on the accretion disc, taking advantage of the relativistic invariant $(I_{\nu}/\nu^3)$. The specific flux will be then,

\be F_{\nu_{obs}}=\int \mathcal{G}^3 I_e(\mathcal{G}^{-1} \nu_{obs}) d\Omega_{obs}, \ee
where $\mathcal{G}=\left(\frac{\nu_{obs}}{\nu_e}\right)$ is a redshift factor. Instead of calculating the specific flux (which depends on the distance of the source) it would be more convenient to calculate the corresponding isotropic luminosity

\begin{figure*}
\includegraphics[width=.35\textwidth]{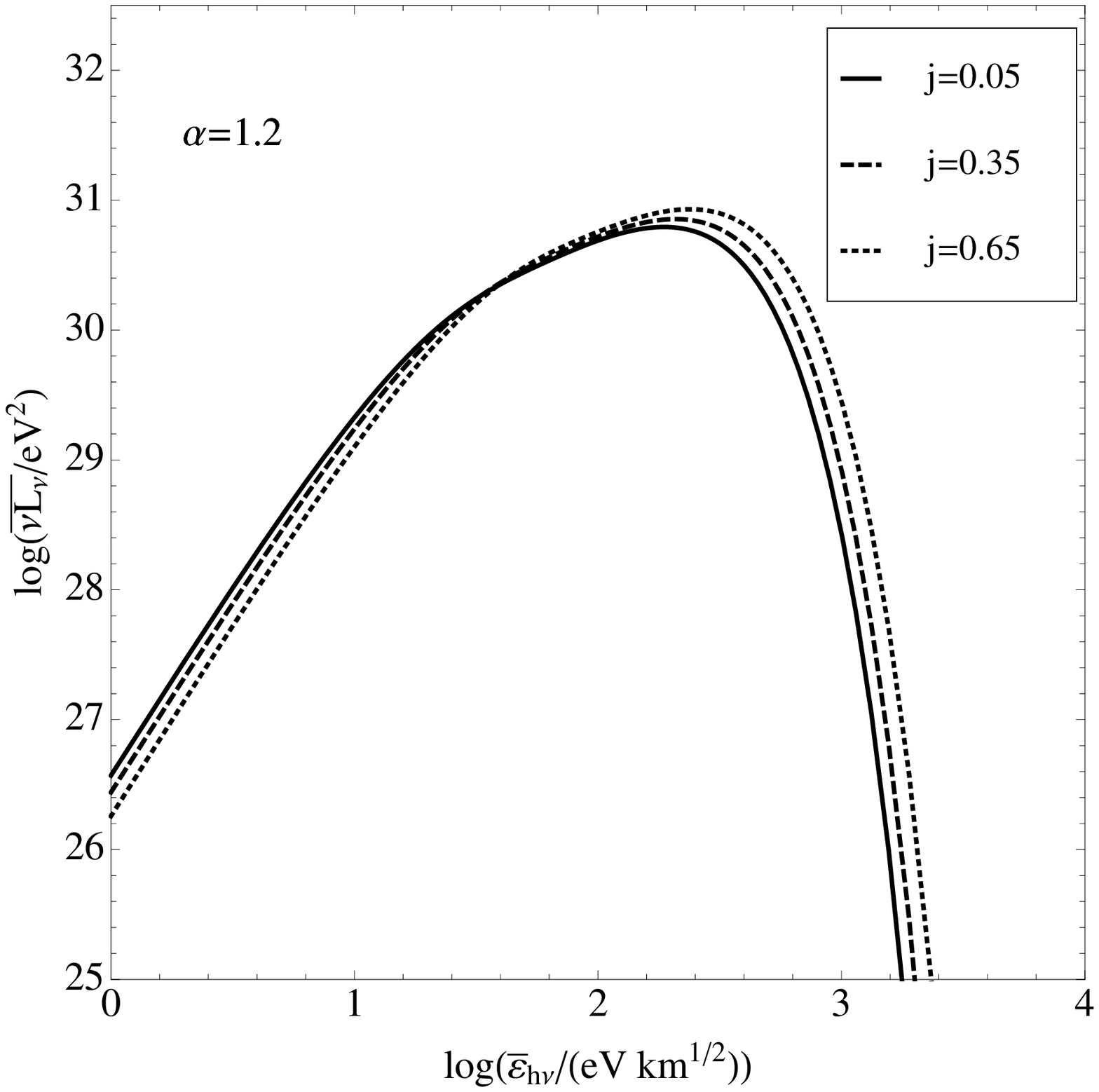} 
\includegraphics[width=.35\textwidth]{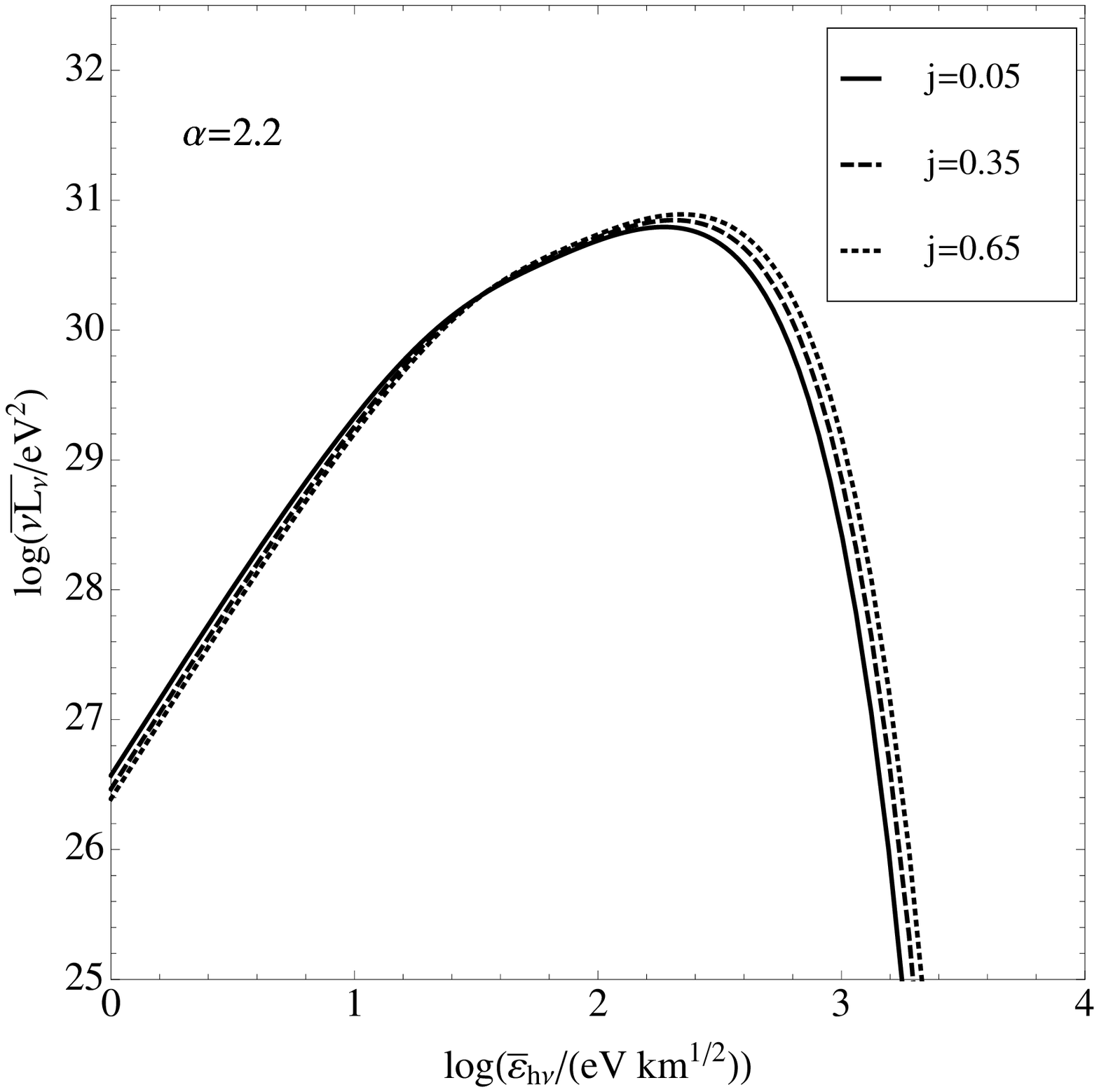}
\includegraphics[width=.35\textwidth]{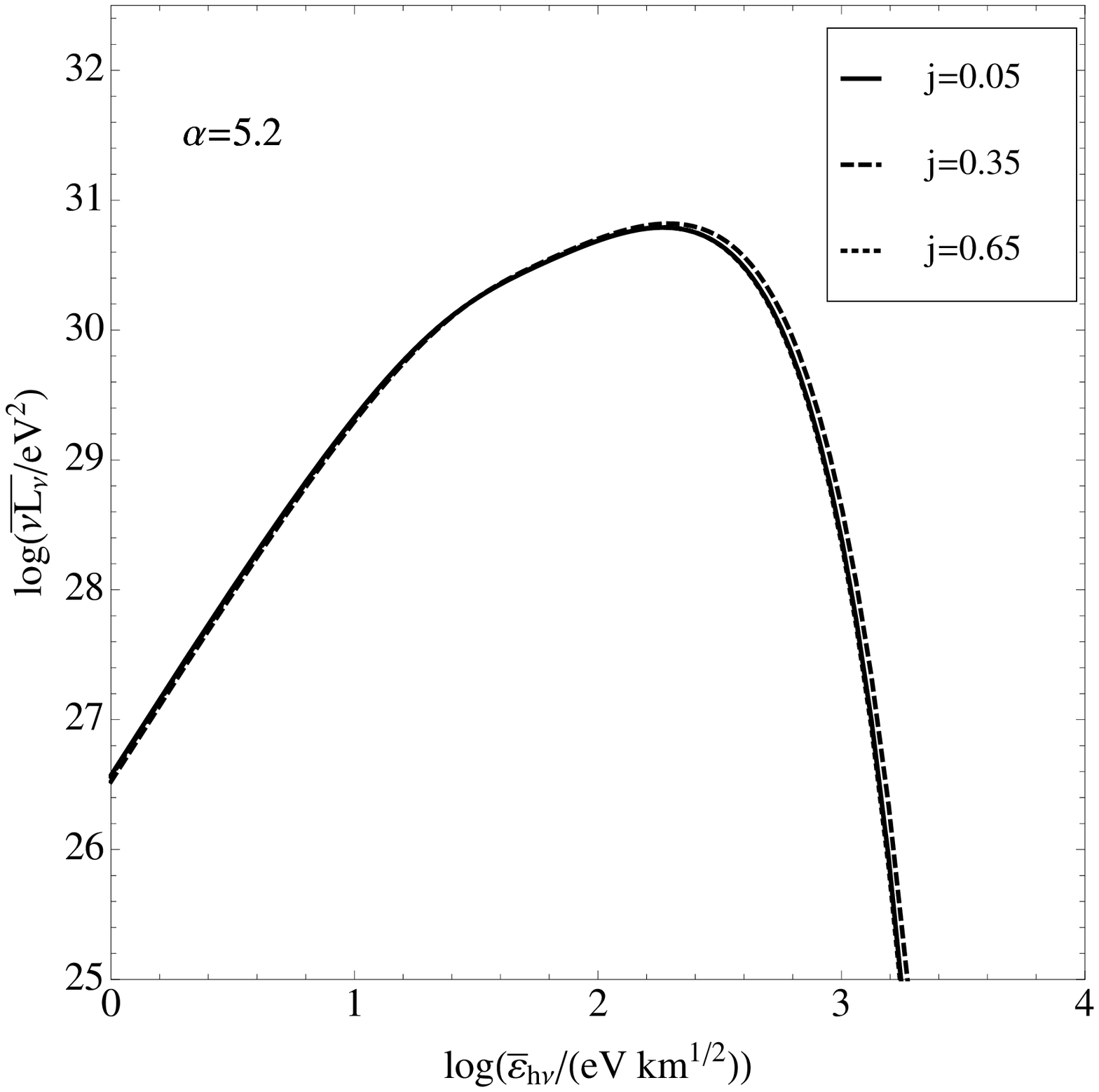} 
\includegraphics[width=.35\textwidth]{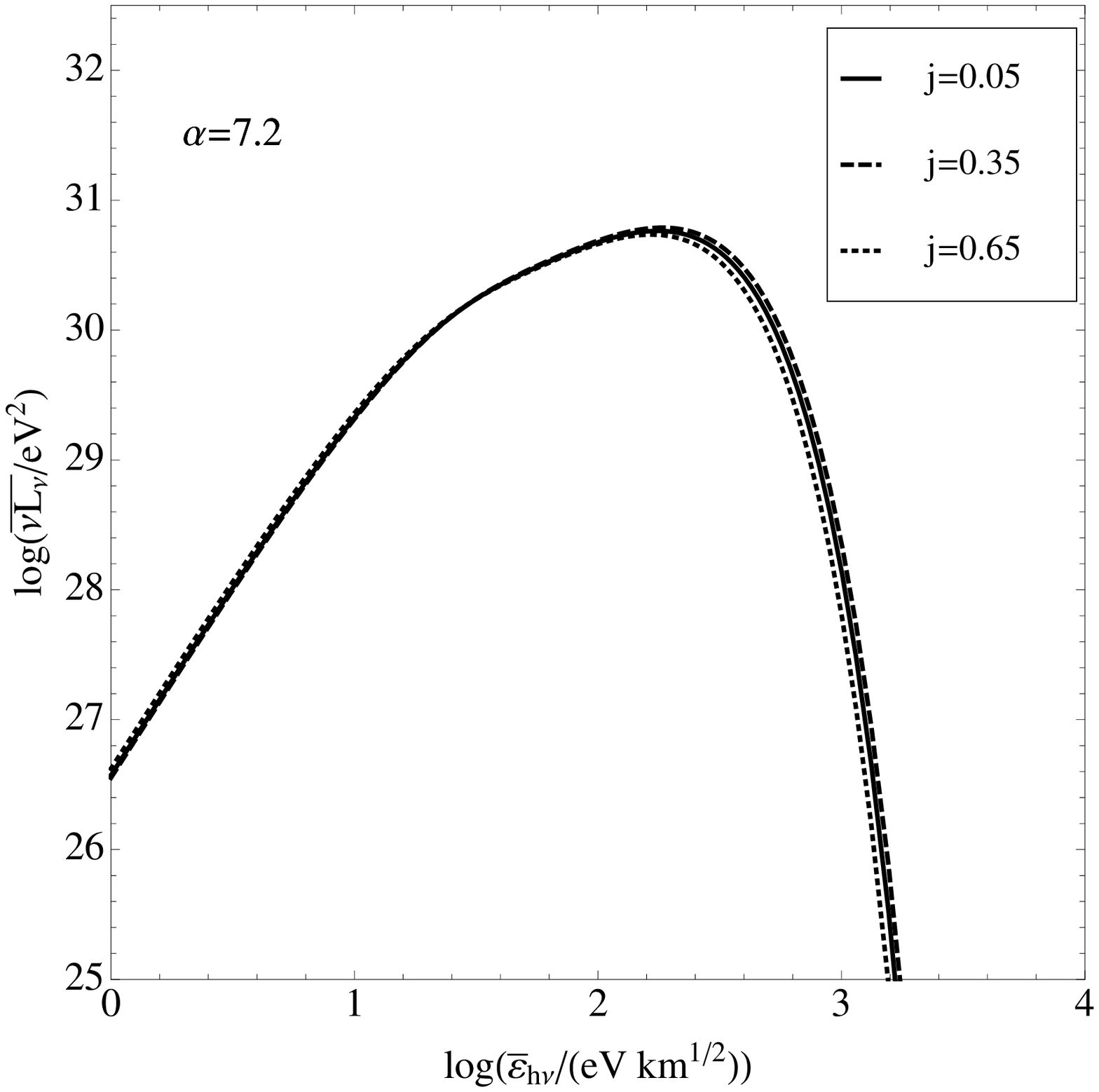}
\caption{Indicative spectra produced for the same models as the ones used to produce the temperature profiles in Fig. \ref{Tprofiles}. The plots show the logarithm of the integrated luminosity $\nu L_{\nu}$ in eV$^2$ for a given accretion rate, against the mass-reduced observed photon energy $\bar{\varepsilon}_{h\nu}$ in eV km$^{1/2}$. The accretion rate used is $\dot{M}_0=10^{-12}M_{\odot} \textrm{year}^{-1}$ and the inclination of the disc with respect to the observer is assumed to be zero (face-on). We should also note that $1\textrm{eV}=1.6\times10^{-12}\textrm{erg}$ and that $1\textrm{eV}/h=2.4\times10^{14}s^{-1}$, where $h$ is Planck's constant.
}
\protect\label{Spectra}
\end{figure*}

\be L_{\nu}=4\pi D^2 F_{\nu}, \ee
where $D$ is the distance of the source from the observer, which in the end is independent of the distance since $d\Omega_{obs}=d\Sigma/D^2$, where $d\Sigma$ is the surface at the source. The final expression for the luminosity in terms of the observed photon energy (neglecting light-bending effects) will be,

\be L_{\varepsilon_{obs}}=8\pi \cos i \int \frac{\varepsilon_{obs}^3}{(hc)^2} \frac{\sqrt{-g}d\rho d\phi}{\left[ \exp\left(\frac{\varepsilon_{obs}}{\mathcal{G}k_B T_{\textrm{eff}}}\right)-1 \right]} , \ee
where $i$ is the inclination angle, which is zero for a face-on disc and is assumed to be small in our approximation. Since we are neglecting light-bending effects, the redshift factor will be (see for example \cite{Bambi2012ApJ})

\be \mathcal{G}=\frac{\sqrt{-g_{tt}-2g_{t\phi}\Omega-g_{\phi\phi}\Omega^2}}{1+\Omega \rho \sin\phi \sin i}, \ee
where $\Omega$ is the orbital frequency of the emitting fluid element on the accretion disc. Therefore, to calculate the spectrum we need to integrate from the inner edge of the disc up to the outer edge. For our calculation, the inner edge will be either the location of the ISCO or the surface, depending on the model, as it was discussed in the previous sections. For the outer edge we will assume a radius of around $10^3M$. 

At this point we should make some comments on how the various quantities scale with the mass of the central object and the accretion rate. As we discussed in section \ref{sec:thinDisc}, the effective temperature scales with the mass and the accretion rate as $T=M^{-1/2}(\dot{M}_0)^{1/4}\bar{T}$. This means that the energy of the emitted and observed photons will also scale as $\varepsilon_{h\nu}=M^{-1/2}(\dot{M}_0)^{1/4}\bar{\varepsilon}_{h\nu}$. If we introduce that to the expression for the luminosity we can see that it scales as $L_{\nu}=M^{1/2}(\dot{M}_0)^{3/4}\bar{L}_{\nu}$, while the integrated luminosity will scale as $\nu L_{\nu}=(\dot{M}_0)\overline{\nu L}_{\nu}$. This latter scaling of course is to be expected, since the total emitted energy rate can only depend on the matter accretion rate, i.e., on the available energy. Therefore, the integrated luminosity can be consider as a measure of the accretion rate (independent of the mass of the central object).

Fig. \ref{Spectra} shows some indicative spectra of accretion discs that are face-on with respect to the observer and extend from some inner radius up to some outer radius. The vertical axis is the logarithm of the integrated luminosity in units of eV$^2$ while the horizontal axis is the logarithm of the mass-reduced energy of the emitted photons as measured by the observer in eV km$^{1/2}$. The models that we have used to construct the spectra are the same as the models that were used to produce the temperature profiles in Fig. \ref{Tprofiles}. The two top plots have spin and quadrupolar deformation that correspond to models which have an ISCO, while for the two bottom plots the disc terminates on the surface of the star. The outer radius of the disc is at around $1000M$. In the spectra, three typical regions stand out. At the low energy side we can see the low energy tail of the blackbody spectrum from the region of the outer radius of the disc, which should scale as $\sim\varepsilon^3$ (the slope in the plot is $\sim3$). At the high energy side, we can see the high energy exponential behaviour of the inner edge of the disc. Finally, between the two we can see a modified blackbody spectrum that follows a power law $\sim\varepsilon^{\alpha}$. The high energy side of the spectrum traces the behaviour of the maximum temperature of the disc, which is higher the deeper in the gravitational potential the inner edge of the disc reaches. Also, the peak of the integrated luminosity is consistent with the assumed accretion rate of $\dot{M}_0=10^{-12}M_{\odot} \textrm{year}^{-1}$.     

\begin{figure}
\centering
\includegraphics[width=.4\textwidth]{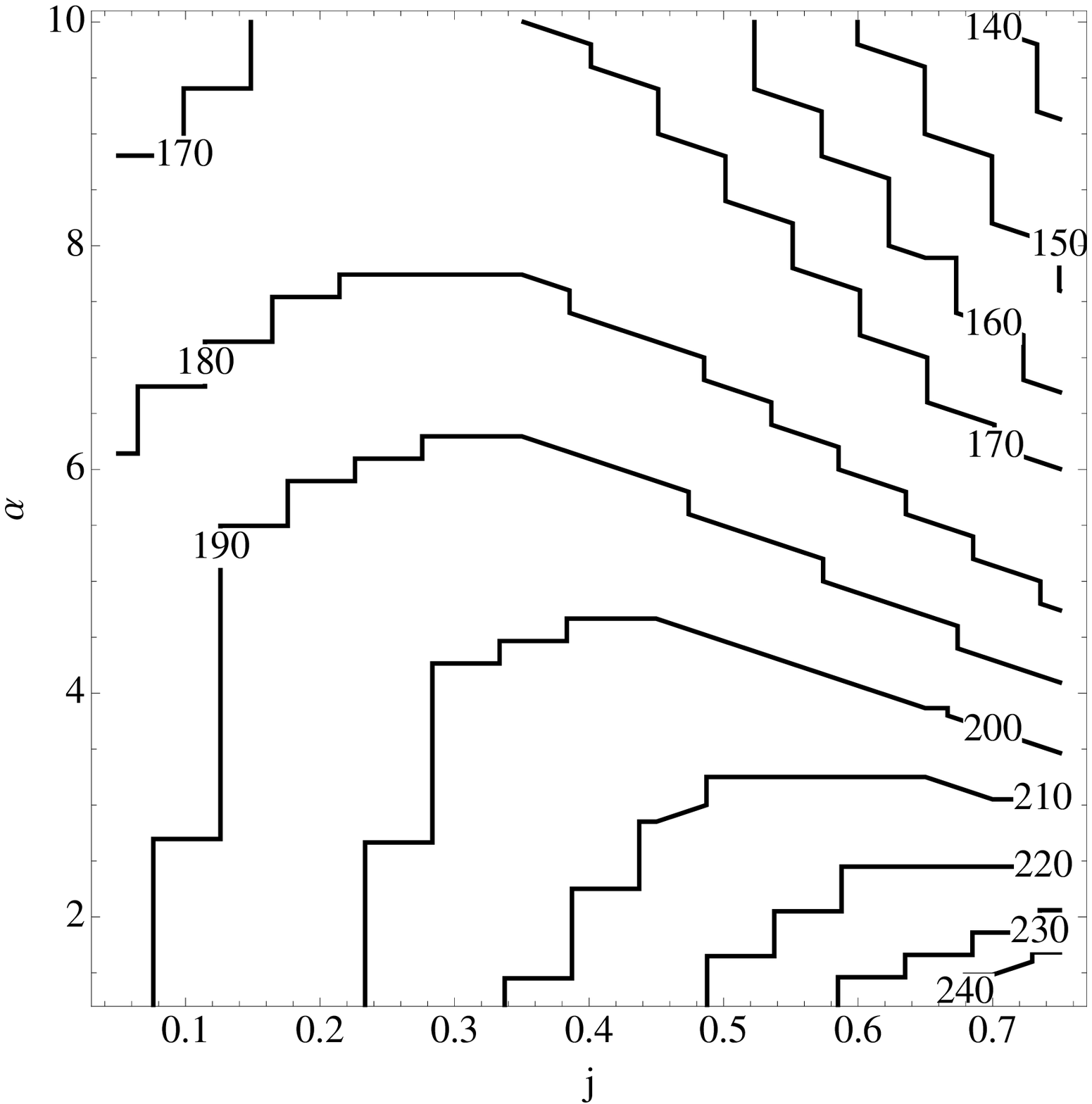}
\caption{Contours of the mass-reduced photon energy in eV km$^{1/2}$ where the maximum energy output is observed in the emitted spectrum (which is essentially the photon energy where we have the maximum of the $\nu L_{\nu}$ spectrum). As before, we assume an accretion rate of $\dot{M}_0=10^{-12}M_{\odot} \textrm{year}^{-1}$.
}
\protect\label{maxphoton}
\end{figure}

\begin{figure}
\centering
\includegraphics[width=.4\textwidth]{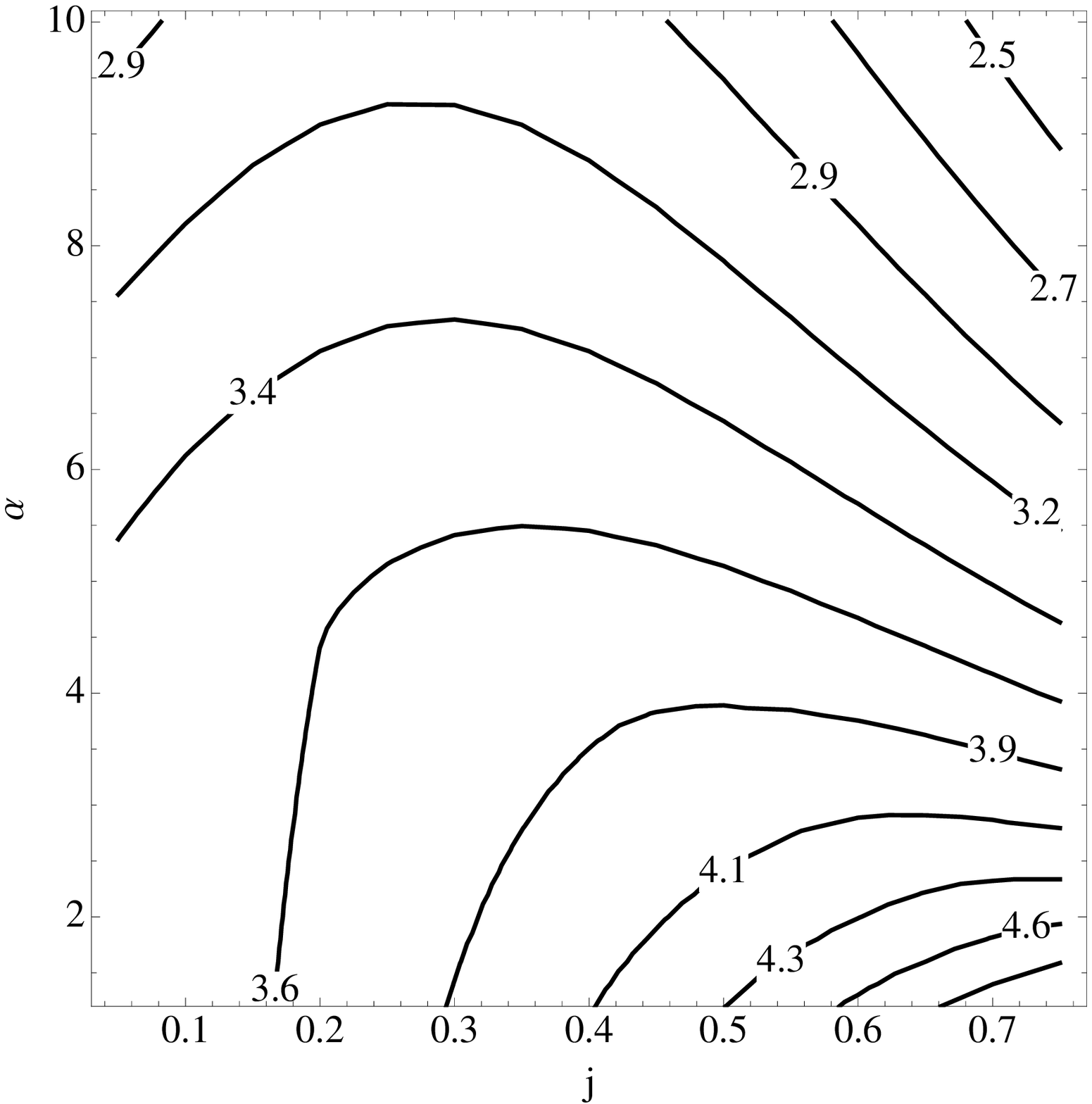}
\caption{Contours of maximum $\nu L_{\nu}/\dot{M}_0$, expressed in percentile, from the spectrum. This quantity is essentially an alternative measure of the efficiency of the accretion.
}
\protect\label{maxL}
\end{figure}

From these spectra we could obtain two observables. The energy of the photons where the maximum of the integrated luminosity is observed and the value of that maximum integrated luminosity. These quantities are given in contour plots in terms of the spin parameter and the quadrupolar deformability in Figs \ref{maxphoton} and \ref{maxL} respectively. 

Specifically, Fig. \ref{maxphoton} gives contours of the mass-reduced energy of the photons at the maximum of the spectrum in units of eV km$^{1/2}$, while Fig. \ref{maxL} gives contours of the maximum integrated luminosity divided by the mass accretion rate, $\nu L_{\nu}/\dot{M}_0$. This last quantity is essentially an alternative measure of the efficiency of the accretion, which we saw in section \ref{sec:efficiency}.

The analysis performed here and the spectra obtained demonstrate that one could have an analytic parametrised description of an accretion disc around a NS in a way that is independent of the EoS and depends only on the NS's spin and quadrupole. Having thus an analytic spacetime that describes the geometry exterior to NSs is a very useful tool. One could further use this spacetime to do more detailed and realistic calculations using ray-tracing and taking into account all the other effects that we have neglected here for simplicity. 

Expressing the various observables of this and the previous sections in terms of a two dimensional parameter space can be quite powerful in probing the properties of NSs in GR. We will try to demonstrate this in the following section.

\section{Combining observables to constrain the EoS}
\label{sec:4}
 
In section \ref{sec:observables} we discussed various possible observables associated with the orbital motion around a NS, in conjunction to some NS properties discussed in the appendix. 
The observables/properties are, the location of the ISCO, $R_{\textrm{ISCO}}$, the location of the surface of the star on the equatorial plane, i.e., the stellar equatorial radius $R_{eq}$, the orbital and the nodal precession frequencies of the innermost available circular orbit, which depending on the model is either the location of the ISCO or the stellar surface, the rotation frequency of the star, the orbital and periastron precession frequencies at the location where the nodal precession frequency becomes zero, the efficiency of a thin accretion disc around the star, the maximum effective temperature of the disc, the photon energy at the maximum of the integrated luminosity and finally the maximum integrated luminosity itself. All these quantities were mass-reduced, therefore for our analysis we have assumed throughout that the mass of the star is independently known.  

But even after having an independent measurement of the mass, not all of these quantities are immediately observable, as is the case for the $R_{\textrm{ISCO}}$ for example. Some of these though can be indirectly related to astrophysical observables. The various frequencies for example could be observable if they are associated with QPOs, as is assumed in the context of the relativistic precession model (\cite{stella}). Correspondingly, measurements of the X-ray spectra of LMXBs could provide the efficiency and the integrated luminosities or the relevant photon energies. Furthermore it might be possible in the future to perform disc tomography and obtain thus the temperature distributions of the discs in LMXB systems. 

\begin{figure*}
\centering
\includegraphics[width=.3\textwidth]{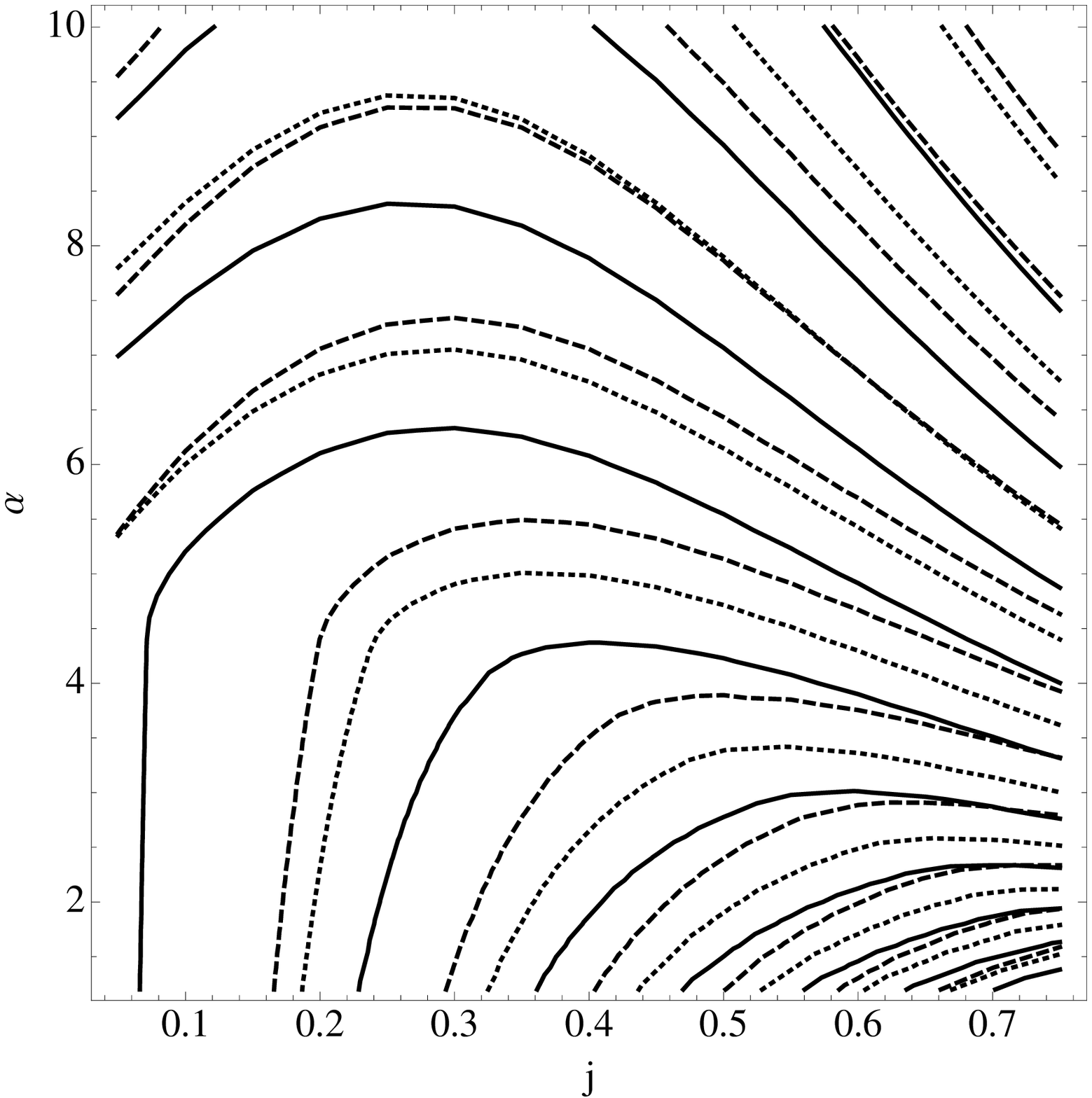}
\includegraphics[width=.3\textwidth]{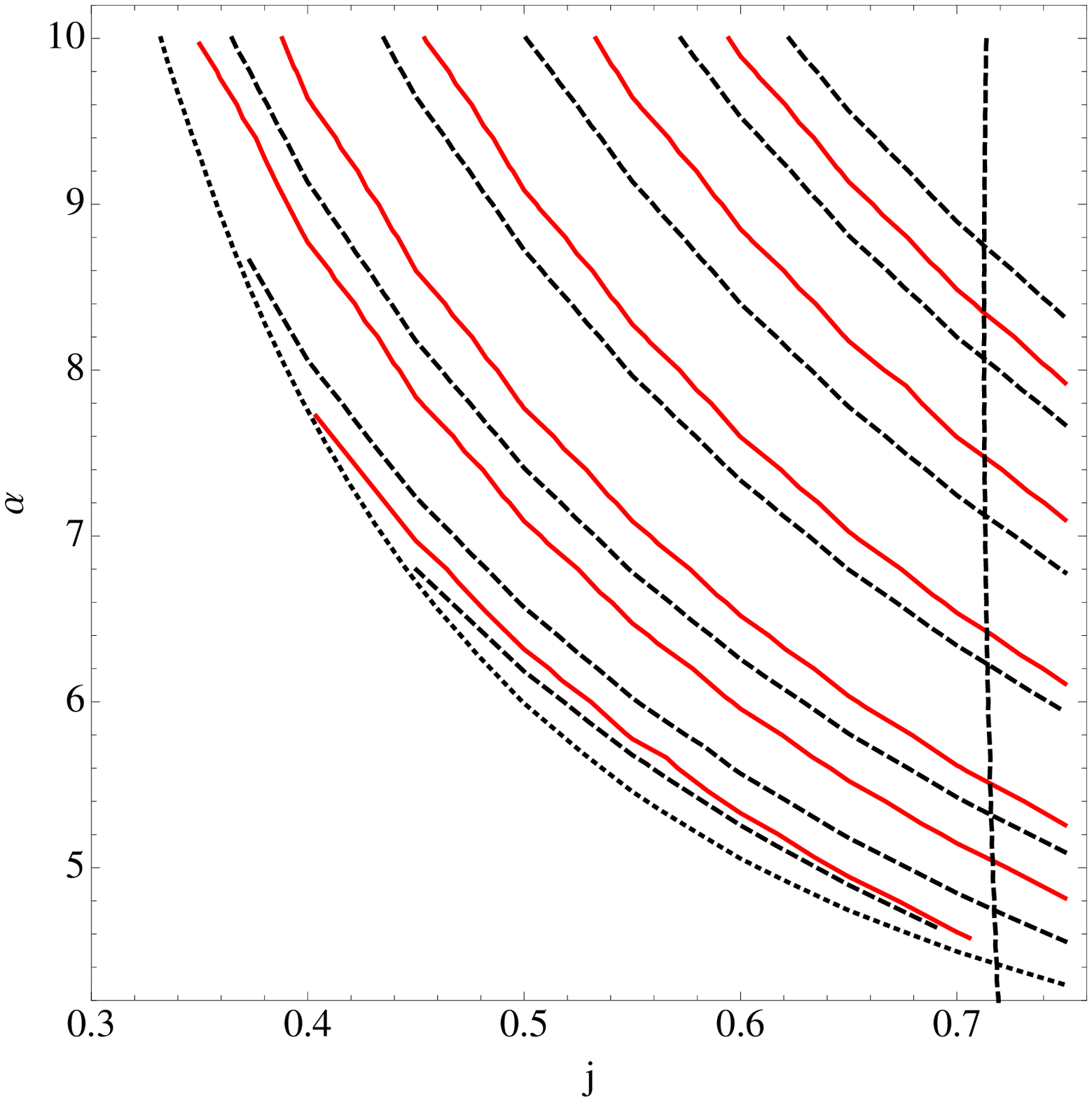}
\includegraphics[width=.3\textwidth]{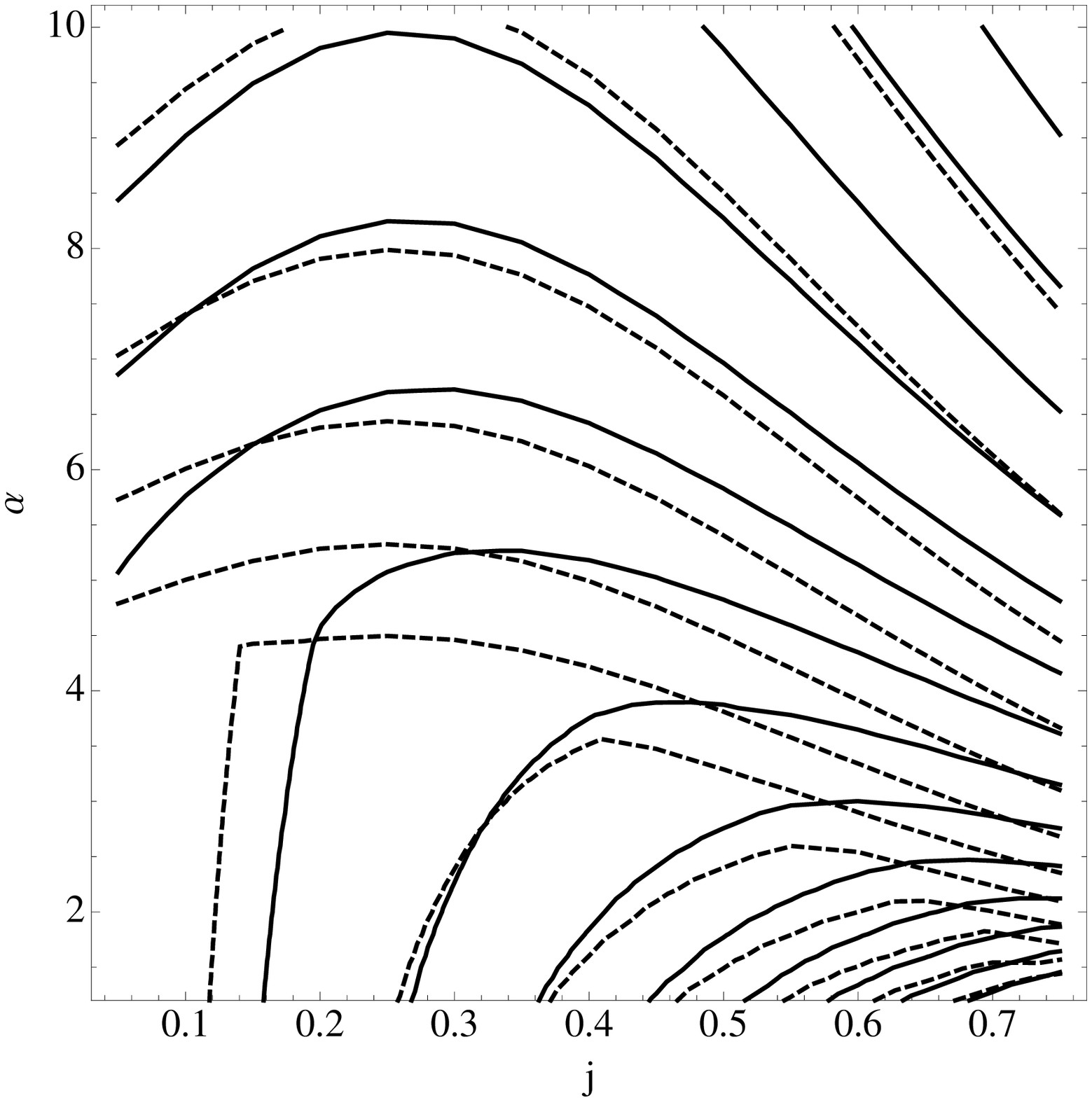}
\caption{Left: Contour plots of the accretion efficiency $\eta$ (dotted lines), the maximum effective temperature of the disc $T_{\textrm{eff}}^{max}$ (solid lines) and the maximum integrated luminosity $\nu L_{\nu}/\dot{M}_0$ (dashed lines). This plot is a combination of the plots in Figs \ref{efficiency}, \ref{maxTeff}, and \ref{maxL}. One can see that these quantities are in a sense conjugate, since their contours are almost parallel. Middle: Plot of the orbital frequency (dashed black curve) and the periastron precession frequency (solid red curve) at the location where the nodal precession becomes zero. This is the same plot as in Fig. \ref{OmegaZisco3}. Right: This plot shows the maximum effective temperature of the disc $T_{\textrm{eff}}^{max}$ (solid lines) and the orbital frequency $\nu_{K}^{max}$ (dashed lines) of the nearest possible stelar orbit. These two quantities are not as clearly conjugate as the quantities in the other plots, but they are clearly not orthogonal. All the quantities plotted and compared are appropriately mass-reduced. The numerical values of the relevant quantities on each curve have been omitted for clarity of presentation, but can be found in the figures of section \ref{sec:observables}. 
}
\protect\label{conjObserv}
\end{figure*}

Observing the figures in section \ref{sec:observables} one can see that there are quantities that we will call here ``conjugate'', i.e., they give contours that seem to be almost parallel, while there are other quantities that we will call ``orthogonal'', i.e., their contours have clear intersections and form a grid. These properties of the observables can be used to perform consistency checks and ``measurements''. The quantities that are conjugate can be used for consistency checks that could distinguish between proper identifications of the astrophysical observables to the corresponding geometric quantities on the one hand and possible misidentifications on the other hand. For example, the observed QPO frequencies could be caused by a different effect than relativistic precession. In that case, the comparison between conjugate quantities could help to realize the misidentification. In Fig. \ref{conjObserv} we present some of these conjugate quantities discussed in the previous section. We present these plots only for illustrative purposes and for that reason as well as for clarity of presentation, we have removed the contour values (which can be found in the figures in the previous section). In the plot on the left of Fig. \ref{conjObserv} we can see the various quantities that are related to the radiation emitted from the accretion disc. The plot in the middle is the same figure as Fig. \ref{OmegaZisco3} and it shows the orbital frequency and the periastron precession frequency at the location where the nodal precession becomes zero. The plot on the right shows the contours of the orbital frequency of the innermost possible orbit, against the contours of the maximum effective temperature of the accretion disc. 

We should emphasize again that these conjugate properties could be used as a selection tool. It is important to make sure that the quantities that we have measured correspond to the theoretical quantities that are calculated here. In a different case, any analysis based on misidentified observables would give results that would not correspond to the properties of the observed system. In the literature of the continuum fitting method, there are some selection criteria that have to do with the assumption of a thin accretion disc. The main requirement is that the luminosity should be $L/L_E \lesssim 0.3$, where $L_E$ is the Eddington luminosity, so that the thin disc assumption does not break down (see the review by \cite{McClintock2014SSRv} and references therein). In our case, we can use these different conjugate quantities as an additional criterion on whether the observed properties of a particular system are those that we assume in the theoretical analysis presented here. For example, if the QPO frequency that we measure from a system is not the orbital frequency of the innermost circular orbit, then we would not expect it to have a correlation to the maximum effective temperature of the accretion disc as the one that we observe in the right plot of Fig. \ref{conjObserv}.

On the other hand, orthogonal quantities can be used to perform measurements of the model parameters. The measurement of two such quantities would give an intersection point in the parameter space of $(j,\alpha)$ and therefore a measurement of the $j$ and $\alpha$ of the particular NS. The additional knowledge of the mass of the NS would imply the knowledge of the first three multipole moments of the star, i.e., the mass $M$, the angular momentum $J \equiv j M^2$ and the quadrupole $Q \equiv M_2 \equiv -\alpha j^2 M^3$. The knowledge of these three parameters for a large enough number of NSs can result in constraints for the EoS (\cite{Pappas:2013naa}). We should note that this approach on calculating the first three multipole moments of the spacetime and of the central compact object, is an alternative to the approach proposed by \cite{PappasQPOs} for fitting the evolution of QPO frequencies, that was further explored by \cite{Boshkayev:2014mua} and \cite{Boshkayev:2015mna}.       

\begin{figure*}
\centering
\includegraphics[width=.4\textwidth]{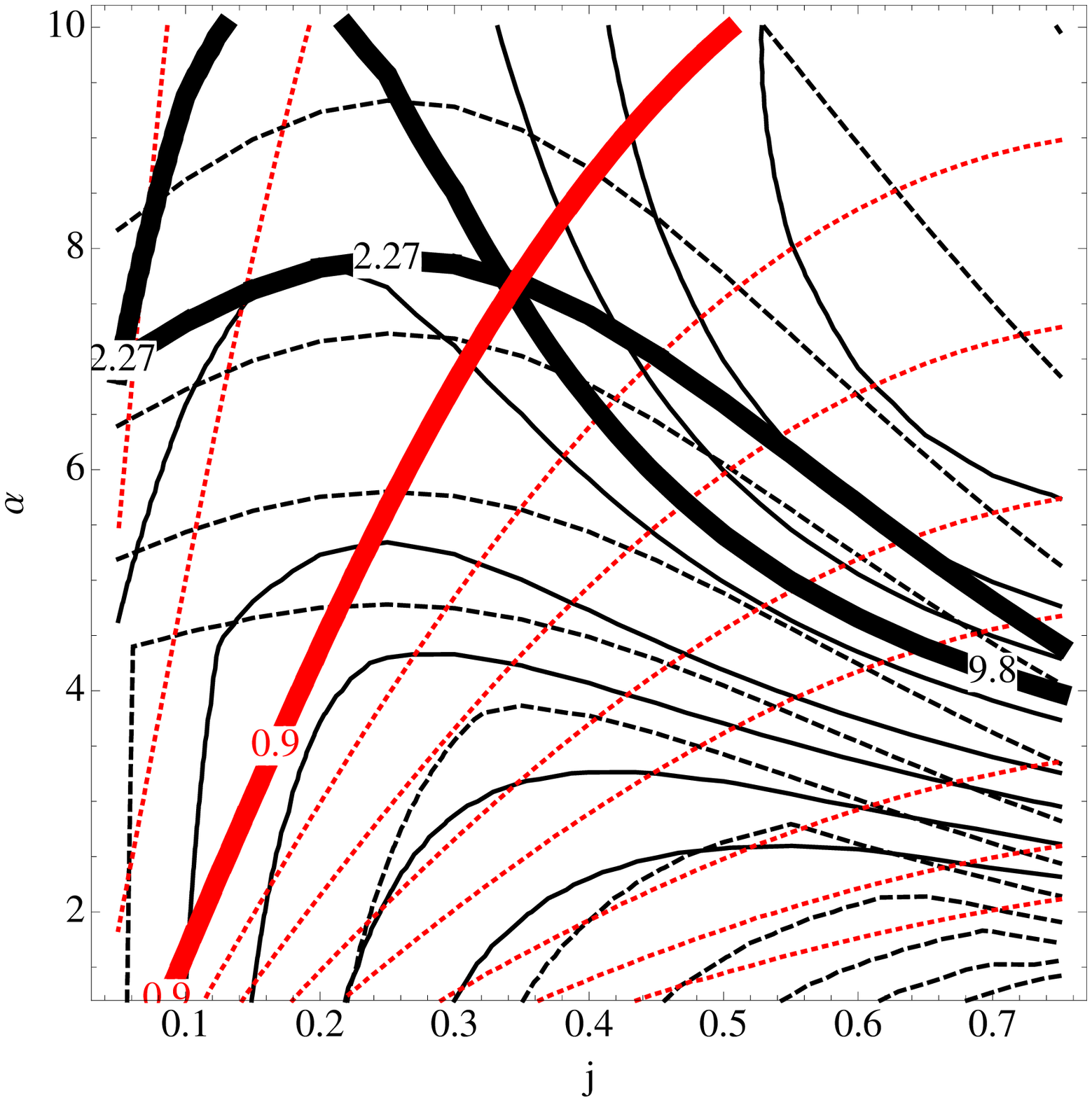}
\includegraphics[width=.4\textwidth]{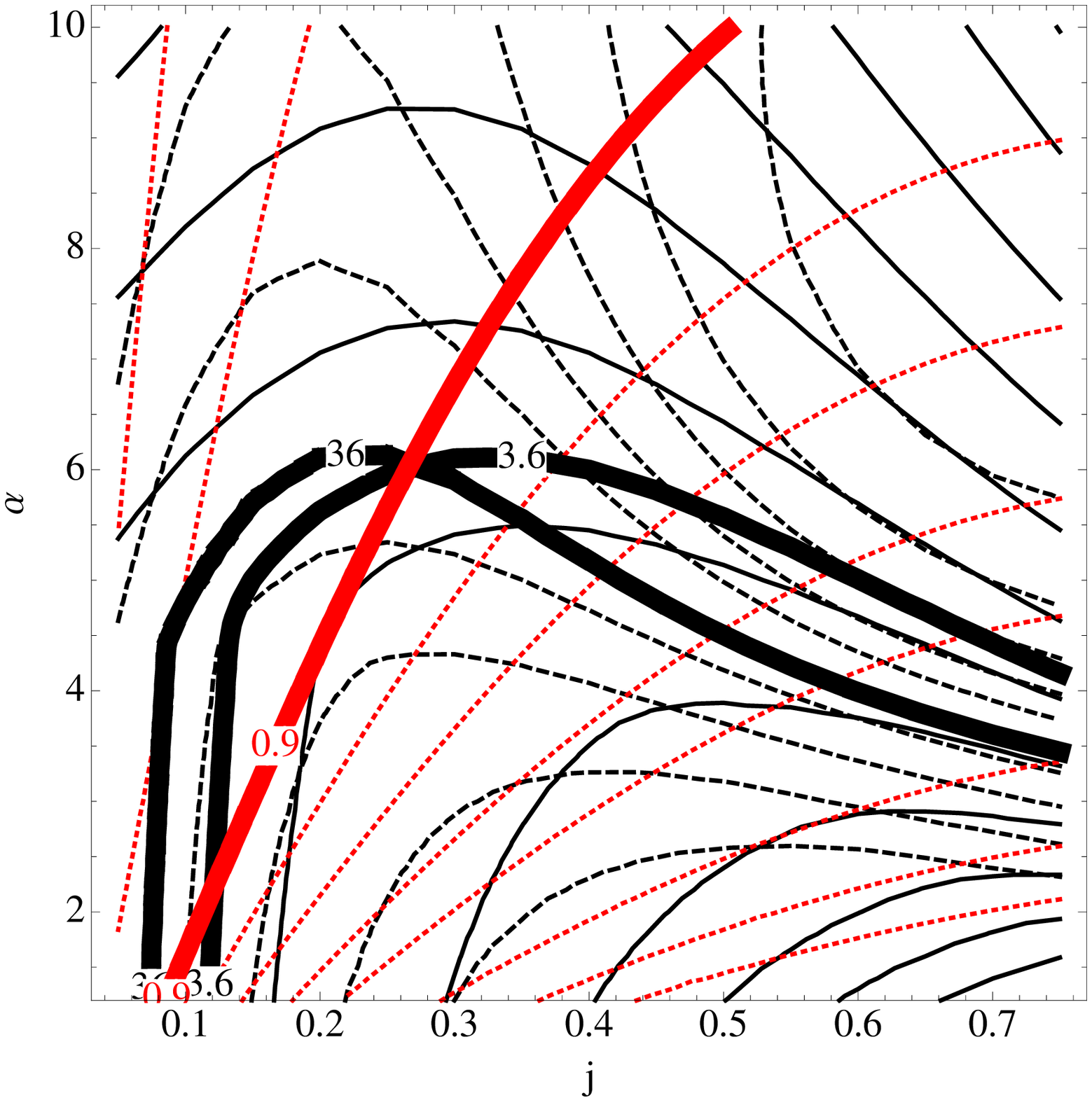}
\caption{Left: Contour plots of the orbital frequency at the orbit closest to the stellar surface (dashed black lines), the nodal precession frequency at the same orbit (solid black lines), and the rotation frequency of the star itself (dotted red lines). Right: Contour plots of the maximum integrated luminosity (solid black lines), the nodal precession frequency at the orbit closest to the stelar surface (dashed black lines), and the rotation frequency of the star itself (dotted red lines). These quantities are orthogonal to each other and allow for the possibility of inferring $j$ and $\alpha$. All the quantities plotted and compared are appropriately mass-reduced. The numerical values of the relevant quantities on each curve have been omitted for clarity of presentation, but can be found in the figures of section \ref{sec:observables}. In each plot we represent with three thick curves the curves that correspond to some measurement of each of the quantities. The intersection of these three curves indicates a measurement of $j$ and $\alpha$.
}
\protect\label{CombObserv}
\end{figure*}

In Fig. \ref{CombObserv} we show some combinations of these orthogonal quantities that could be used to perform measurements. The plot on the left shows contours of the orbital and the nodal precession frequencies of the innermost orbit around the star and of the rotational frequency of the star. If for a particular system we were to identify for example three QPO frequencies that would correspond to this triplet, then the intersection point of the three corresponding curves would indicate a pair of $(j,\alpha)$. An alternative combination is presented in the plot on the right, where we show contours of the nodal precession frequency of the innermost orbit around the star, contours of the rotational frequency of the star and contours of the maximum integrated luminosity. Again we can see that a possible measurement of these three quantities could suggest a pair of $(j,\alpha)$. 

The choice of the quantities plotted in Fig. \ref{CombObserv} is only one possible combination. For example, one could choose to plot the orbital or the precession frequencies at different locations, depending on the preferred model for the QPO frequencies (for an alternative, see for example the work by \cite{Stuchlik2015arXiv}). This demonstrates the power of having a parameterized analytic description of the spacetime, since one could have as an extra free parameter the location at which a QPO is created (assuming the relativistic precession model for example) and try to fit for that parameter as well. 

Another alternative application could be the following, if one were to have two well identified quantities, such as the stellar rotation frequency and the orbital frequency at the innermost circular orbit for example, then one could attempt to test a hypothesis about the nature of a third frequency by seeing if the right curve passes through the point were the two other curves intersect. This would be an even stronger test of the assumptions that enter the modelling of the various astrophysical observables, than the tests one can do using the conjugate quantities.

\section{Conclusions}
\label{sec:conclusions}

In the current work we have performed an investigation of the astrophysical properties of NSs that are related to the spacetime geometry around them. These properties were presented separated in two categories, the first being the properties of the geodesic motion of a test particle or a fluid element around the NS and the second being the properties of matter accreting onto the NS while emitting thermal radiation. 

To calculate the properties of the geodesic orbits and of the accretion disc we have used the analytic two-soliton spacetime, which was shown by \cite{twosoliton} to accurately reproduce the spacetime around NSs if one chooses the parameters of the two-soliton spacetime in such a way so that the spacetime has the same first four non-zero multipole moments as the corresponding NS. Furthermore, instead of assigning specific values for the multipole moments of the NS, we have taken advantage of the recently found universal relations between the multipole moments (see work by \cite{ Pappas:2013naa} and \cite{YagietalM4}). By further mass-reducing all the observable quantities we have managed to give a description of the various orbital and accretion properties in an EoS-independent way that depends on only two parameters, the spin parameter of the NS, $j$, and the quadrupolar deformability, $\alpha$. 

Since the various observables are parameterized only by the spin parameter and the quadrupolar deformability, one can construct contour plots for these observables on the parameter space of $(j,\alpha)$. The utility of these contour plots is twofold. The first is that one could use these plots to test the assumptions that enter the modelling of the various astrophysical observables, such as that QPOs originate from geodesic motion for example or assumptions about the properties of accretion discs. The second is that by combining the right observables, one could in principle measure the spin parameter and the quadrupolar deformability (from the intersection of the curves of two or more observables that have been measured). This application assumes the independent knowledge of the mass of the NS. Therefore once there exists an independent mass measurement, the measurement of $j$ and $\alpha$ implies the knowledge of the first three multipole moments of the NS, i.e., the mass $M$, the angular momentum $J=jM^2$, and the quadrupole moment $Q=-\alpha j^2M^3$. 
The measurement of the first three moments of a variety of NSs will be able to constrain the EoS for the matter inside NSs.      

Furthermore, measuring $j$ and $\alpha$ will give us information for other properties of NSs that are not directly measurable, such as the radius of the NS (see appendix \ref{radii}), which enter the calculations related to other astrophysical processes such as photospheric radius expansions. This would provide the possibility of cross-checking and refining our models. 

All these possibilities, and possibly even more than those this author can conceive, are available to us because the analytic two-soliton spacetime and the universal relations between the multipole moments allow for an analytic parametric description of the exterior of NSs in an EoS-independent way with a minimal set of parameters. The results presented here could be extended further by extending the analysis and the modelling of the accretion disc to include more phenomenology, such as for example taking into account temperature colour corrections or ray tracing for the photons and so on and so forth.    

Finally, this analysis could probably be extended to quark stars as well. On the one hand, it is likely that the two-soliton spacetime could describe the exterior of quark stars, if the right value for the multipole moments is chosen (although this would need to be demonstrated). On the other hand, the universal relations for the multipole moments, used here for NSs, were shown by \cite{YagietalM4} to hold for quark stars as well (with a slight modification). Therefore, one could perform a similar analysis to include quark stars. The difference with respect to the analysis presented here would be on the behaviour of the surface of quark stars, which one would expect it to be at smaller radii than it is for NSs. This means that maybe the surface of a quark star would not interfere with the corresponding ISCO, which would simplify the analysis.

\section*{Acknowledgments}

GP would like to thank Kostas Glampedakis, Hector O. Silva and E. Berti for their useful comments and suggestions on the manuscript. Special thanks to Leo C. Stein for many useful comments and suggestions that helped improve the presentation of this work.
GP has received financial support from the European Research Council under the European Union's Seventh Framework Programme (FP7/2007-2013) / ERC grant agreement n. 306425 ``Challenging General Relativity''.


\bibliographystyle{mn} 
\bibliography{mn-jour,mybibliography}

\appendix
%
\section[]{Equatorial radii of NSs}
\label{radii}

The presentation in this work is based on the fact that it is possible to have an EoS-independent description of the spacetime around a NS. Some of the results presented here also depend on whether the radius of the star is such that it does not hide the region of interest. Therefore it would be useful if we could have an EoS-independent parameterization of the radius of a NS (in particular the reduced circumferential radius $R_{\textrm{circ}}/M$) in terms of the spin parameter $j$ and the quadrupolar deformability parameter $\alpha$. It turns out that such a parameterization is possible. 

\begin{figure}
\centering
\includegraphics[width=.23\textwidth]{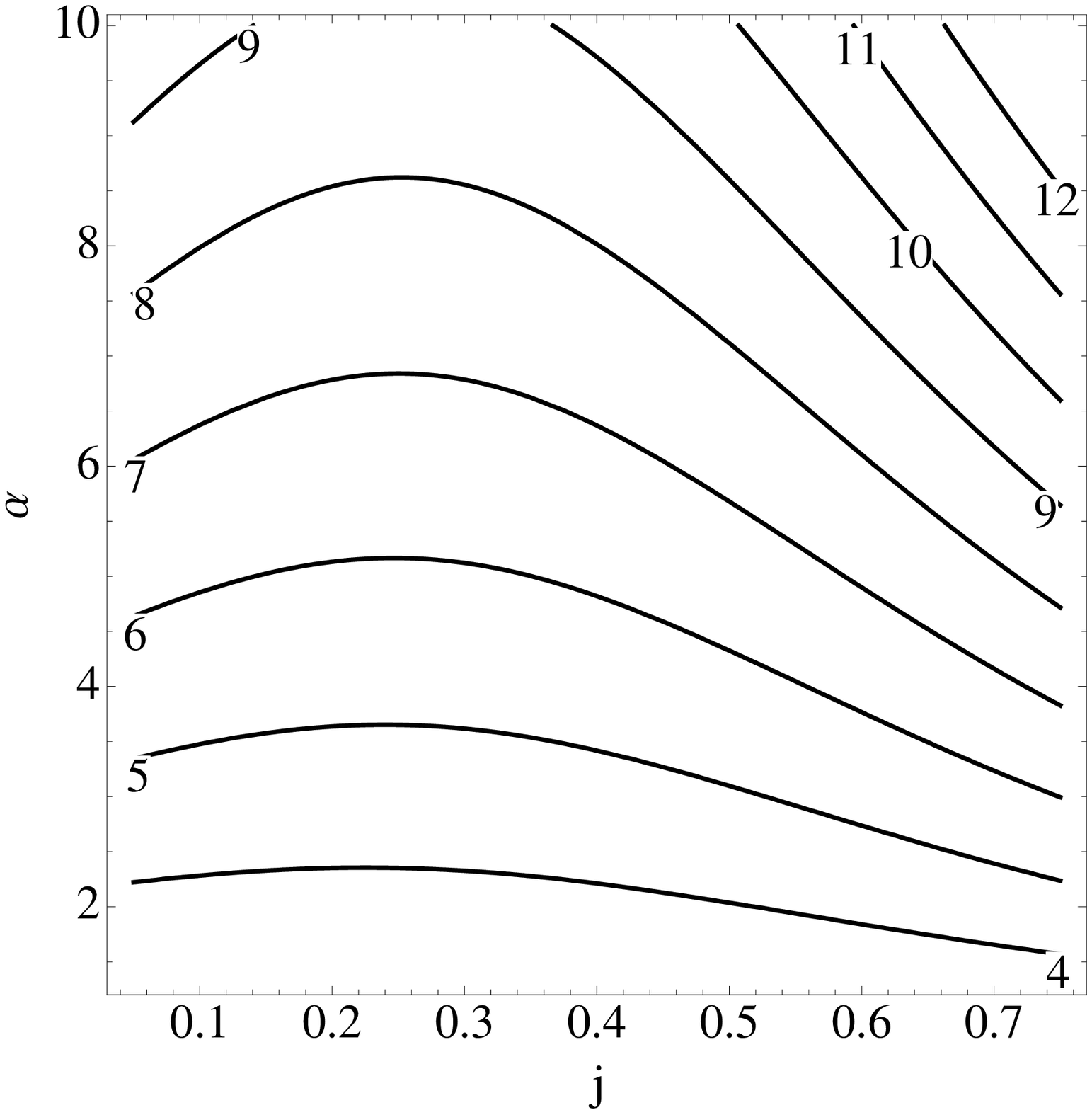} 
\includegraphics[width=.23\textwidth]{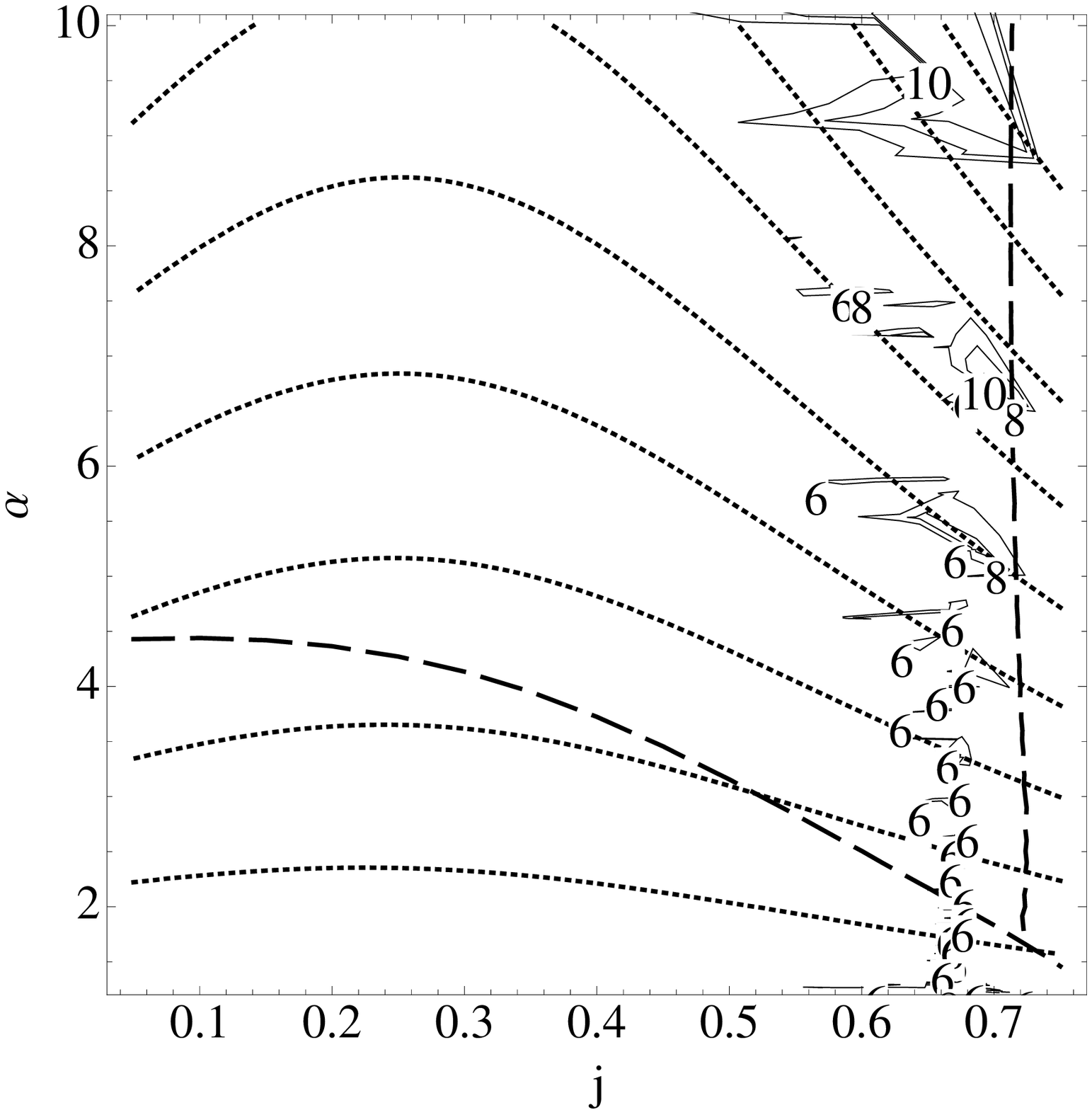} 
\caption{The plot on the left shows contours of the fit of the normalized circumferential equatorial radius $R_{\mathrm{circ}}/M$ of NSs as a function of the spin parameter $j$ and the  quadrupolar deformation parameter $\alpha$. The plot on the right shows overlaid to the normalized radius (dotted contours), contours of relative difference per cent between the actual radius and the fit (solid contours). One can see that the accuracy of the fit is almost for the entire parameter space better than 6 per cent with a vertical region on the side close to maximal rotation that is between 6 and 10 per cent.}
\protect\label{surfacecontour}
\end{figure}

From the set of NS models that were constructed numerically using realistic EoSs in the work by \cite{Pappas:2013naa} (using the RNS numerical code developed by \cite{Sterg}), we plotted the reduced radii as functions of the spin parameter $j$ and the square root of the quadrupolar deformability parameter $\sqrt{\alpha}$. All the models from all the EoSs appear to approximately occupy the same simple surface. Therefore we have fitted the surface with a function of the form,  
\bea R_{eq}/M &=&\mathcal{B}_0+\mathcal{B}_1 j+\mathcal{B}_2 j^2+(\mathcal{A}_0+\mathcal{A}_1 j+\mathcal{A}_2 j^2) (\sqrt{\alpha})^{\mathcal{N}_1}\nn\\
&&+(\mathcal{C}_0+\mathcal{C}_1 j+\mathcal{C}_2 j^2) (\sqrt{\alpha})^{\mathcal{N}_2}, \eea
where the values of the parameters are, $\mathcal{A}_0 = 0.00927584$, $\mathcal{A}_1 = -0.0252801$, $\mathcal{A}_2 = 0.0497335$, $\mathcal{B}_0 = -0.358824$, $\mathcal{B}_1 = 3.15892$, $\mathcal{B}_2 = -5.30171$, $\mathcal{C}_0 = 2.94923$, $\mathcal{C}_1 = -3.20369$, $ \mathcal{C}_2 = 6.02522$, $\mathcal{N}_1 = 4.12566$, and $\mathcal{N}_2 = 0.996284$. For these values, the accuracy with which the surface reproduces the radii of the NSs is better than 10 per cent, with most of the parameter space being described with accuracy better than 6 per cent. It is possible that with a better choice of a fitting function better accuracy can be achieved, but for the purposes of this work the current accuracy will be adequate. The resulting distribution of NS equatorial radii with respect to $j$ and $\alpha$ can be seen in the plot on the left of Fig. \ref{surfacecontour}. On the right we have for illustrative purposes a contour plot of the relative difference between the fit and the actual radii values. The contours indicate the regions where the error is between 6 and 10 per cent.  

This parameterization of the equatorial radius could be also seen in the spirit of reducing the number of free parameters in describing NSs, as it was the aim of the work by \cite{2013Baubocketal}. There has been a similar attempt in expressing the NS radii by \cite{Chakrabarti2014PRL} in an EoS-independent way, the difference being that the radius was given as a function of the spin parameter and the rotation frequency of the star. 

\section[]{Rotation frequency of NSs}
\label{NSrotation}

\begin{figure}
\centering
\includegraphics[width=.23\textwidth]{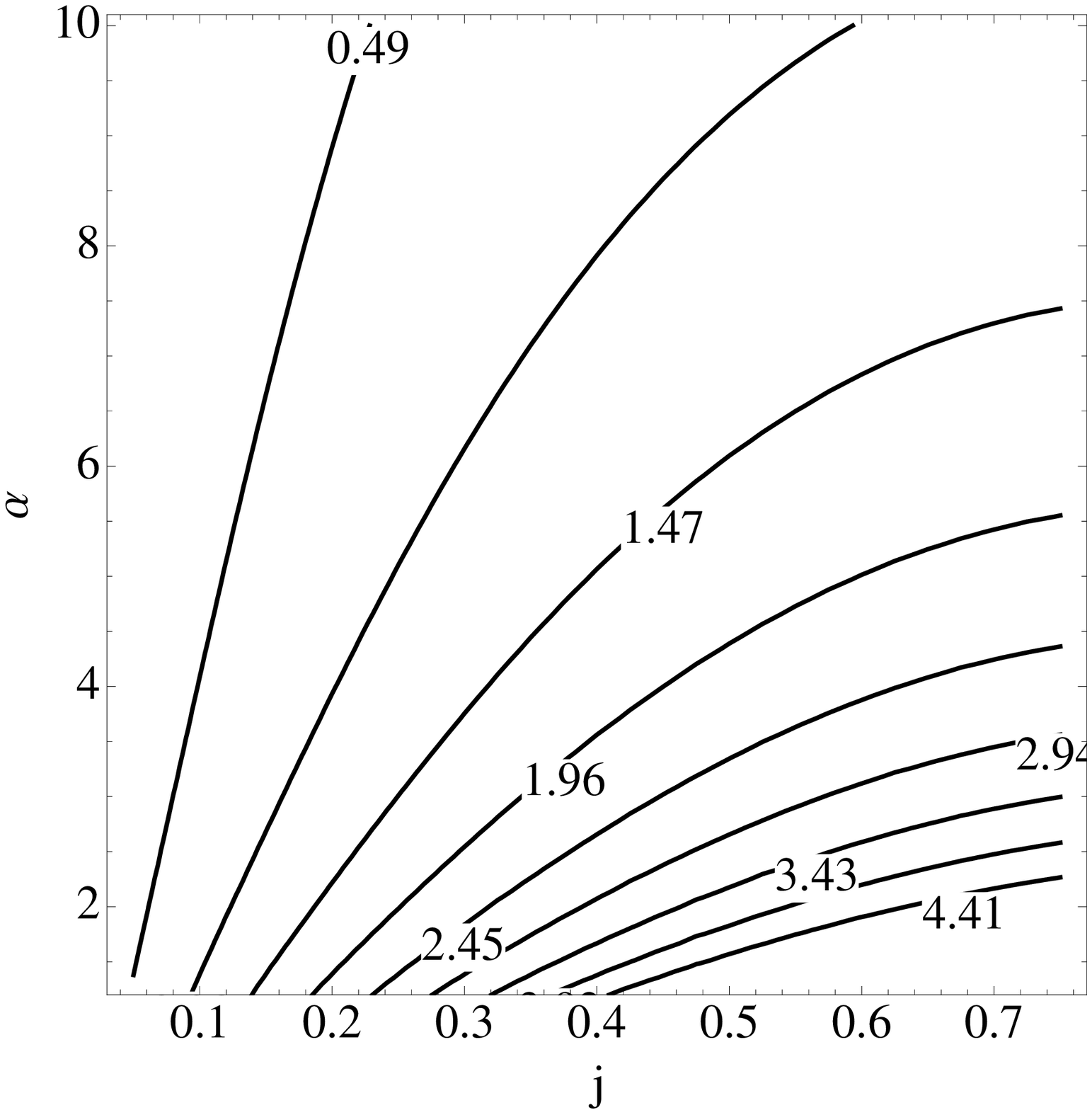} 
\includegraphics[width=.23\textwidth]{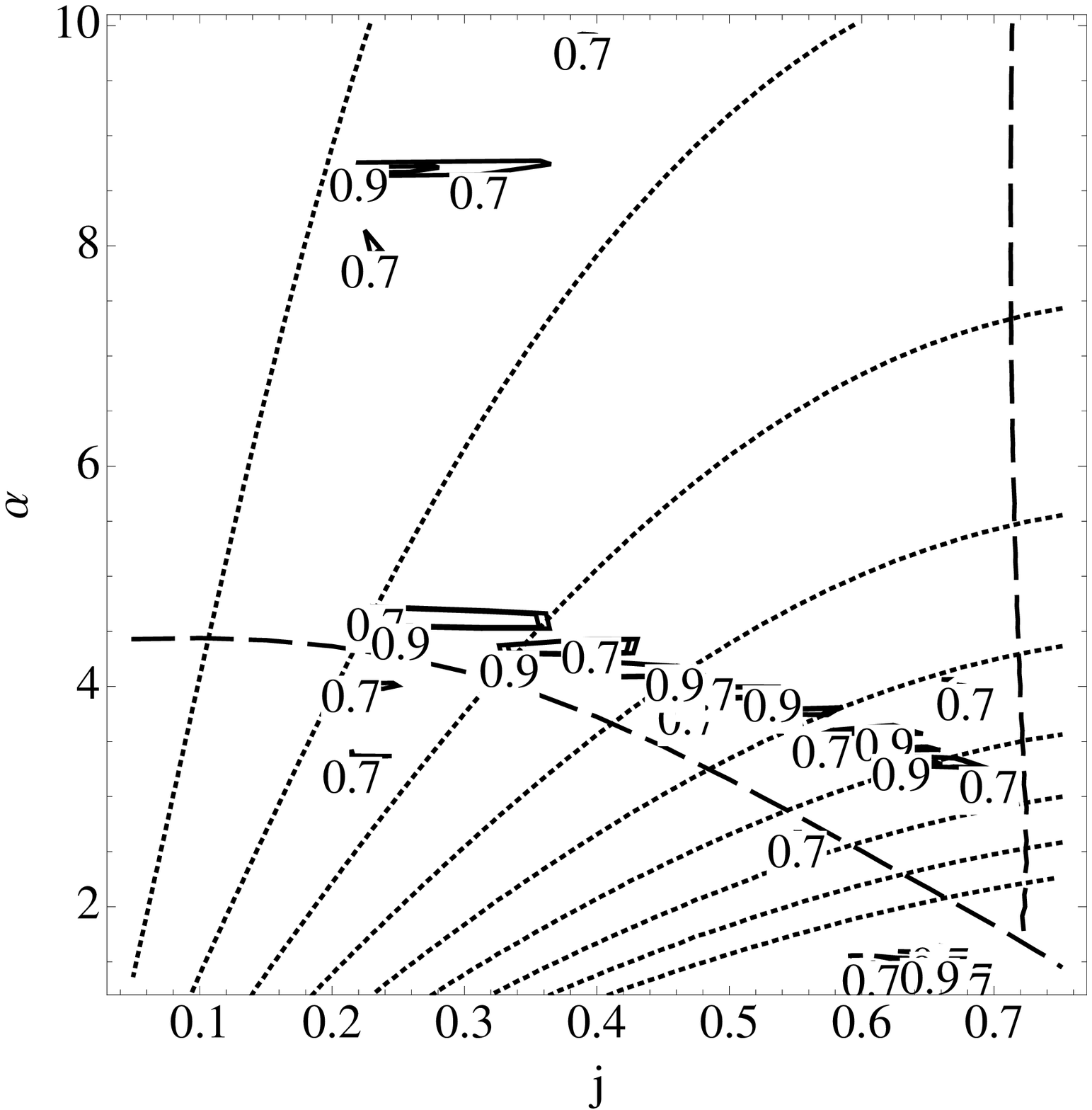} 
\caption{Same figure as Fig. \ref{surfacecontour} but for the stellar rotation frequency. The plot on the left shows contours of the normalized rotation frequency $M\times f (\textrm{km}\times \textrm{kHz})$ of NSs as a function of the spin parameter $j$ and the  quadrupolar deformation parameter $\alpha$. The plot on the right shows overlaid to the normalized rotation frequency (dotted contours), contours of relative difference per cent between the actual rotation and the fit (solid contours). One can see that the accuracy of the fit is for the entire parameter space better than 1 per cent. The plot shows only contours of relative difference larger than 0.7 per cent.}
\protect\label{surfaceRotcontour}
\end{figure}
%
As it was necessary to have an EoS-independent description of the circumferential radius of the star,  it is also necessary to have an EoS-independent parameterization of the stellar rotation rate in terms of the spin parameter $j$ and the quadrupolar deformability parameter $\alpha$. As for the case of the surface radius, from the set of NS models that were constructed numerically using realistic EoSs in the work by \cite{Pappas:2013naa}, we plotted the normalized frequency $M\times f/j$, where $M$ is the mass of the star, $f$ is the rotation frequency, and $j$ is the spin parameter, as functions of the spin parameter $j$ and the square root of the quadrupolar deformability parameter $\sqrt{\alpha}$. All the models from all the EoSs appear to approximately occupy the same simple surface and we have fitted the surface with the same function as the one we used for the radius, which is of the form,  
\bea \frac{M\times f}{j}\!\!\! &=& \!\!\!\mathfrak{B}_0+\mathfrak{B}_1 j+\mathfrak{B}_2 j^2+(\mathfrak{A}_0+\mathfrak{A}_1 j+\mathfrak{A}_2 j^2) (\sqrt{\alpha})^{\mathfrak{n}_1}\nn\\
\!\!\!&&\!\!\!+(\mathfrak{C}_0+\mathfrak{C}_1 j+\mathfrak{C}_2 j^2) (\sqrt{\alpha})^{\mathfrak{n}_2}. \eea
The values of the parameters of the fit are, $\mathfrak{A}_0 = 15.0297$, $\mathfrak{A}_1 = -0.114154$, $\mathfrak{A}_2 = -7.72439$, $\mathfrak{B}_0 = -1.48338$, $\mathfrak{B}_1 = -1.07874$, $\mathfrak{B}_2 = 1.64592$, $\mathfrak{C}_0 = -2.45303$, $\mathfrak{C}_1 = 3.40995$, $\mathfrak{C}_2 = 7.39354$, $\mathfrak{n}_1 =  -1.1698$, and $\mathfrak{n}_2 = -4.50216$. For these values for the fit parameters, the accuracy with which the surface reproduces the normalized frequency $M\times f/j$ of the NSs is everywhere better than 2 per cent and in particular, in the region of the parameter space that is of interest to our analysis, it is better than 1 per cent. The resulting distribution of NS equatorial radii with respect to $j$ and $\alpha$ can be seen in Fig. \ref{surfaceRotcontour}.  

One could note here that $\frac{M\times f}{j}\propto 1/\bar{I}$, where $\bar{I}$ is the reduced moment of inertia as defined by \cite{YY2013Sci,YY2013PhRvD}.

\section[]{Nodal precession frequency $\Omega_z$ around NSs}
\label{app:Nodal}

\begin{figure*}
\includegraphics[width=.39\textwidth]{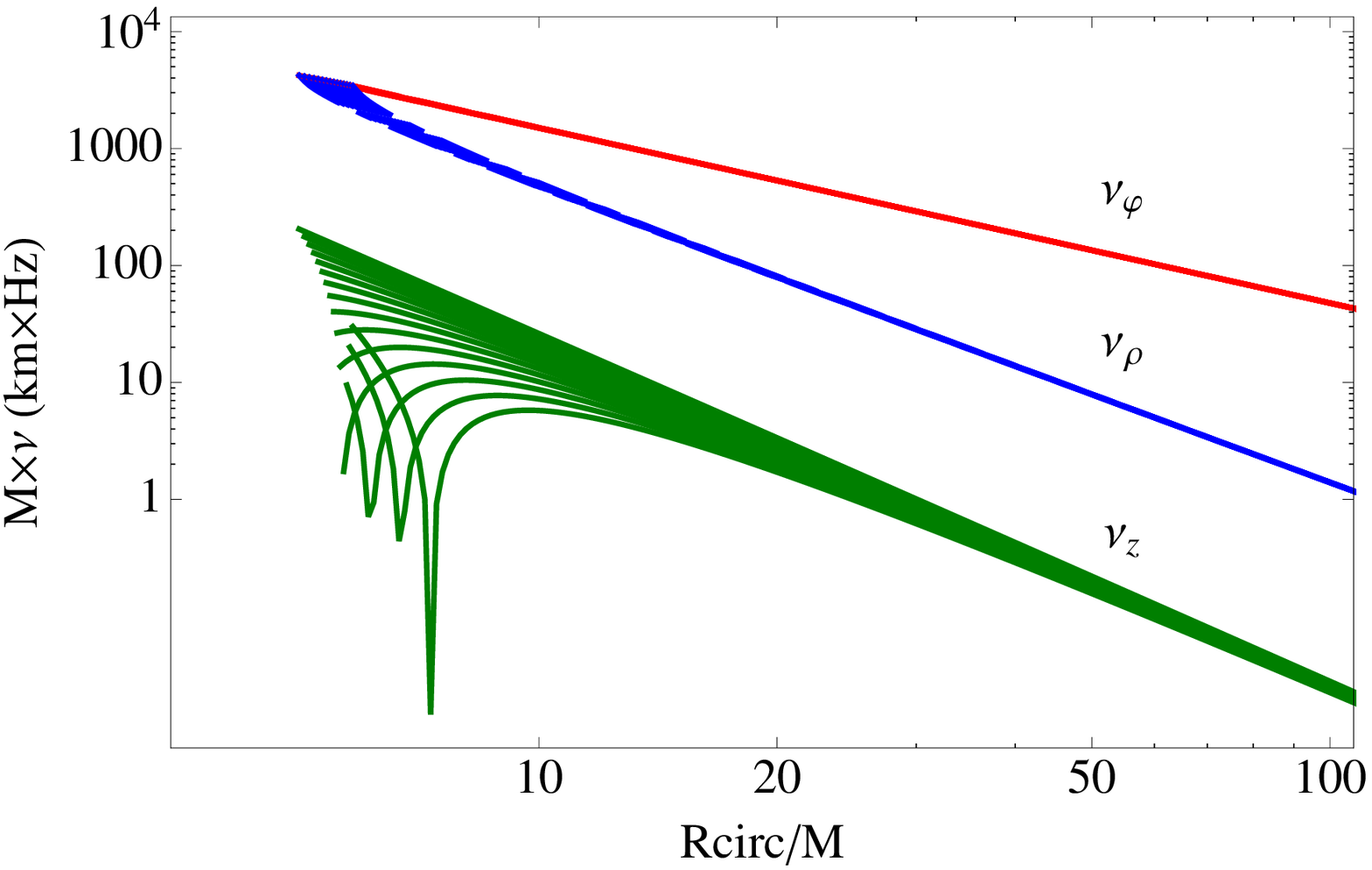} 
\includegraphics[width=.39\textwidth]{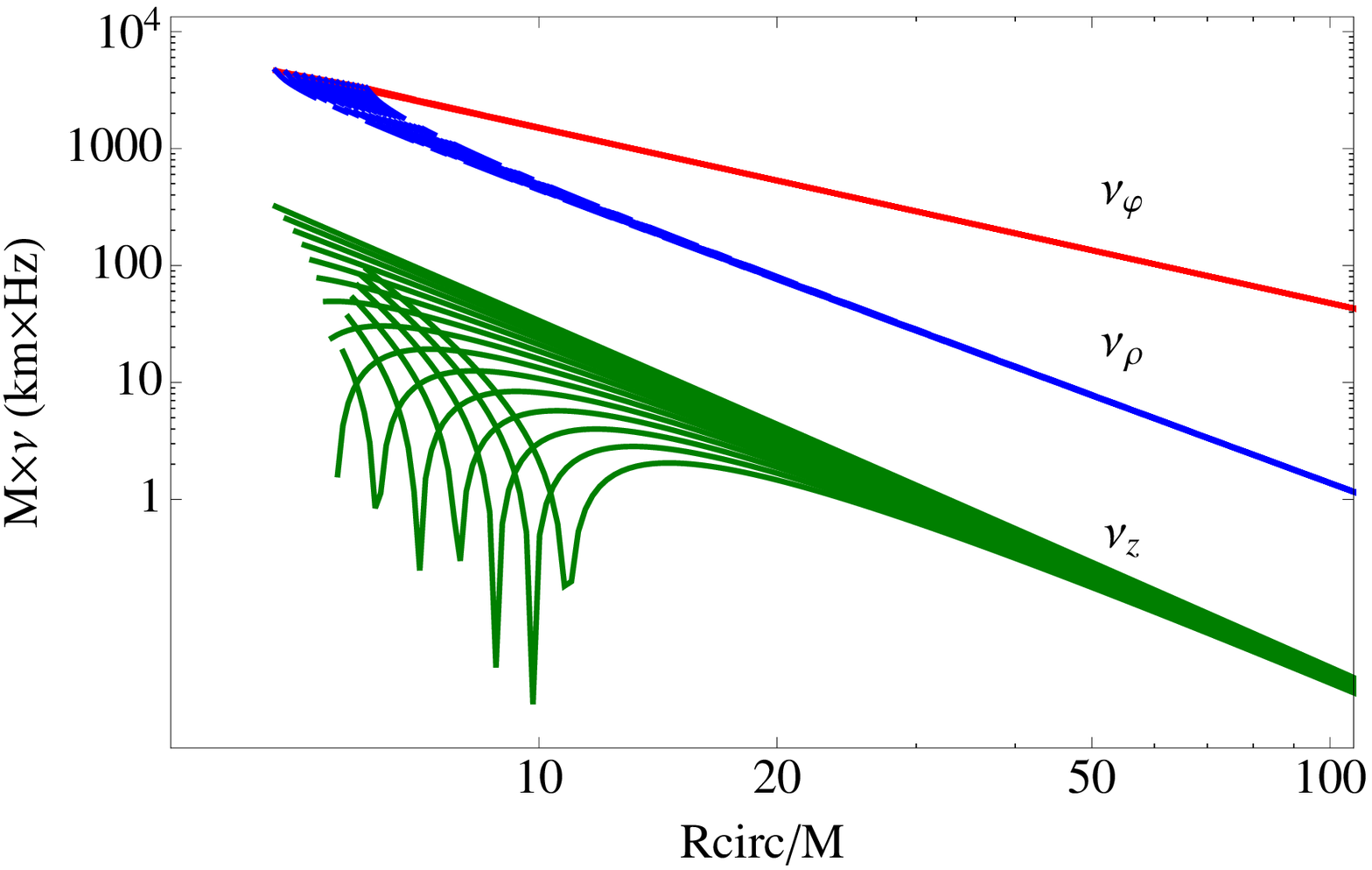}
\includegraphics[width=.39\textwidth]{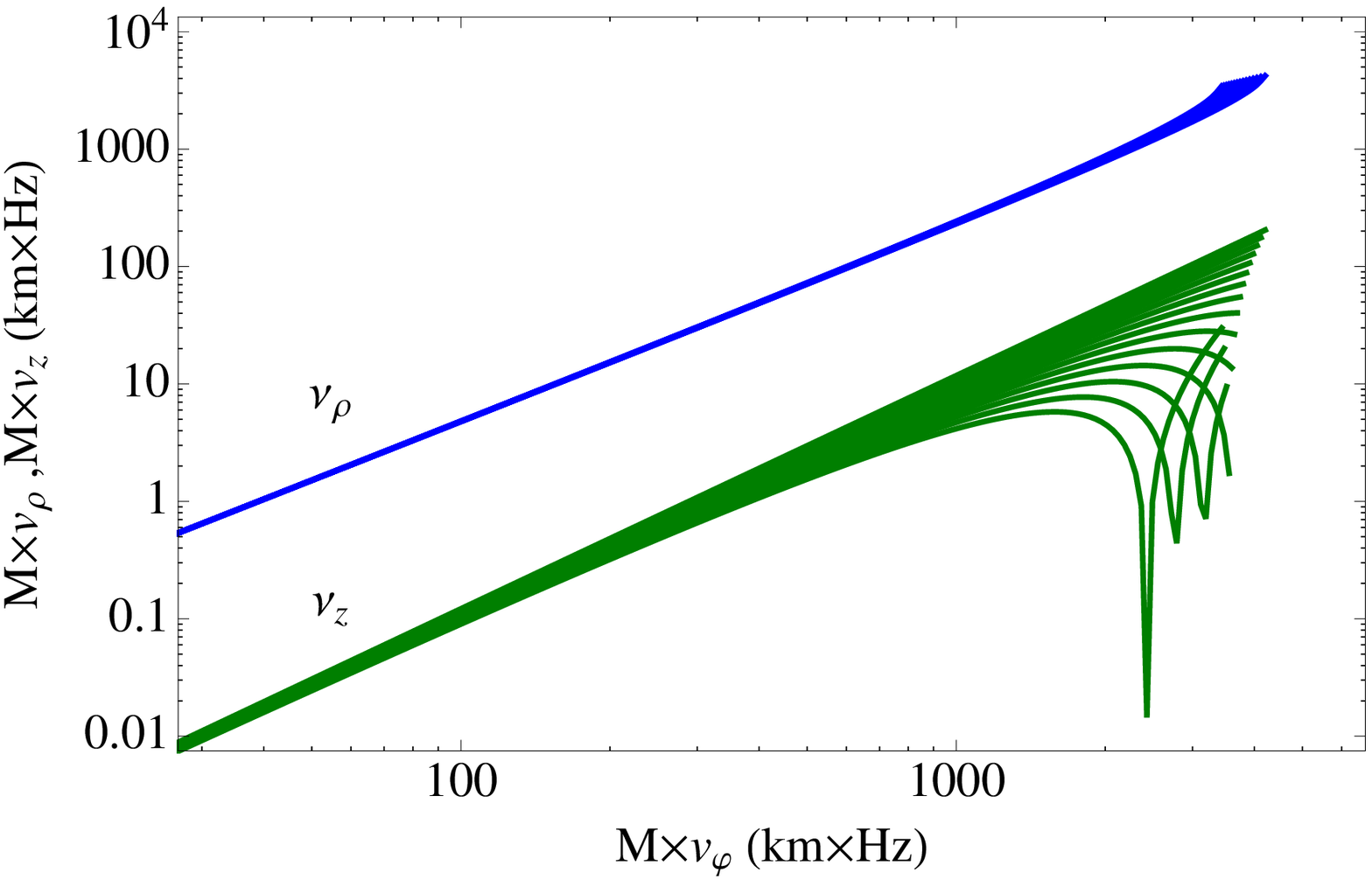} 
\includegraphics[width=.39\textwidth]{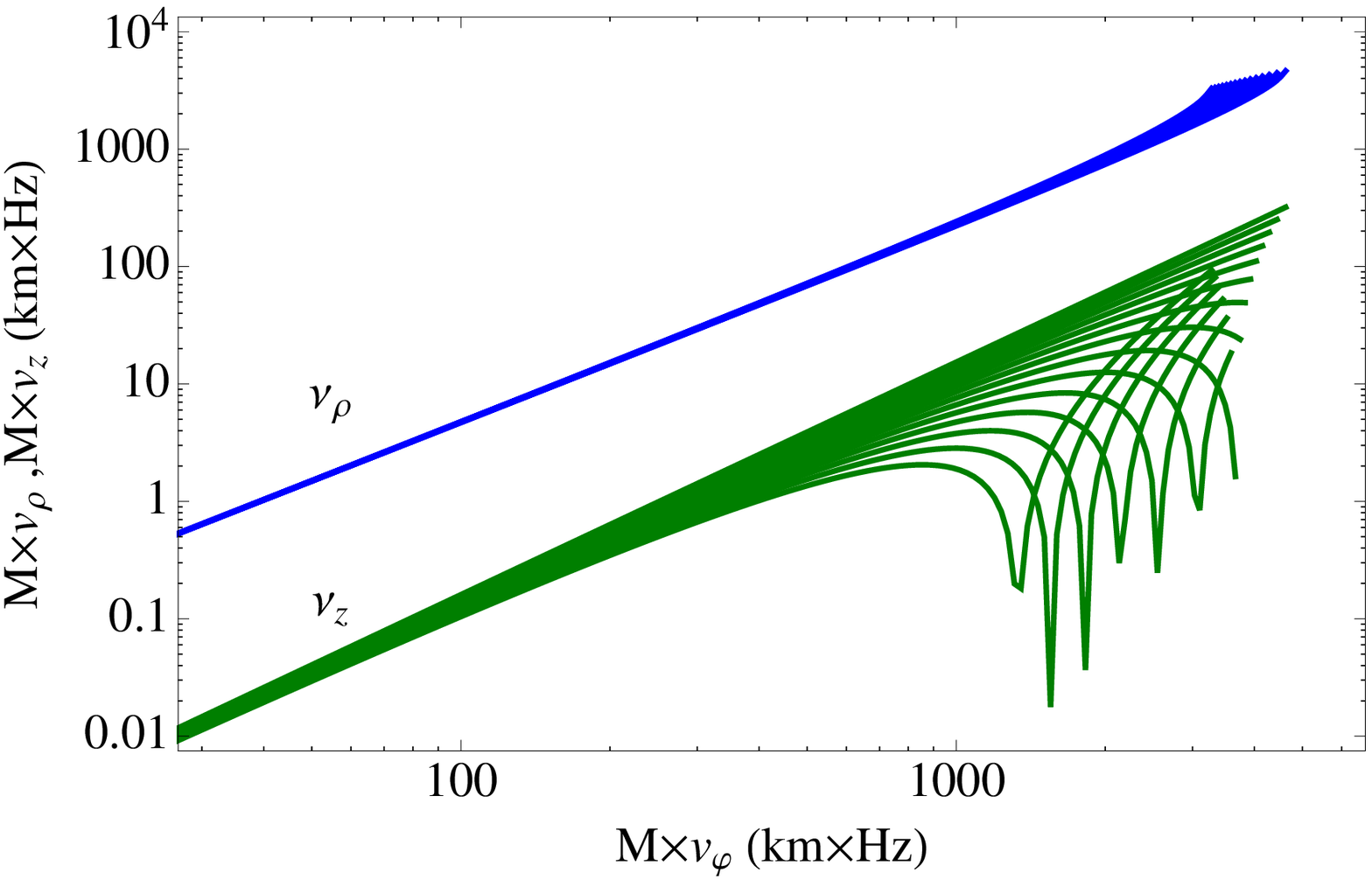}
\caption{The figure shows the various geodesic normalized frequencies for NS spacetimes. The two top panels show the orbital frequency $M\nu_{\varphi}$ (red), the periastron precession frequency $M\nu_{\rho}$ (blue), and the nodal precession frequency $M\nu_z$ (green) as functions of the normalized circumferential distance $R_{\mathrm{circ}}/M$ from the central object. In the case of the nodal precession, since in some regions it becomes negative, we have plotted the absolute value. The two bottom panels show the two precession frequencies (same corresponding colours) as functions of the orbital frequency. The spikes in the plots of the nodal precession indicate where the precession becomes zero. The figures in the left column correspond to compact objects with the same value of the spin parameter $j=0.3$, while the figures on the right column have $j=0.4$. The different curves in each figure correspond to different values of the quadrupolar deformability in the range $1.2\leq \alpha \leq 10$.}
\protect\label{Nodalprecession}
\end{figure*}

\begin{figure*}
\includegraphics[width=.39\textwidth]{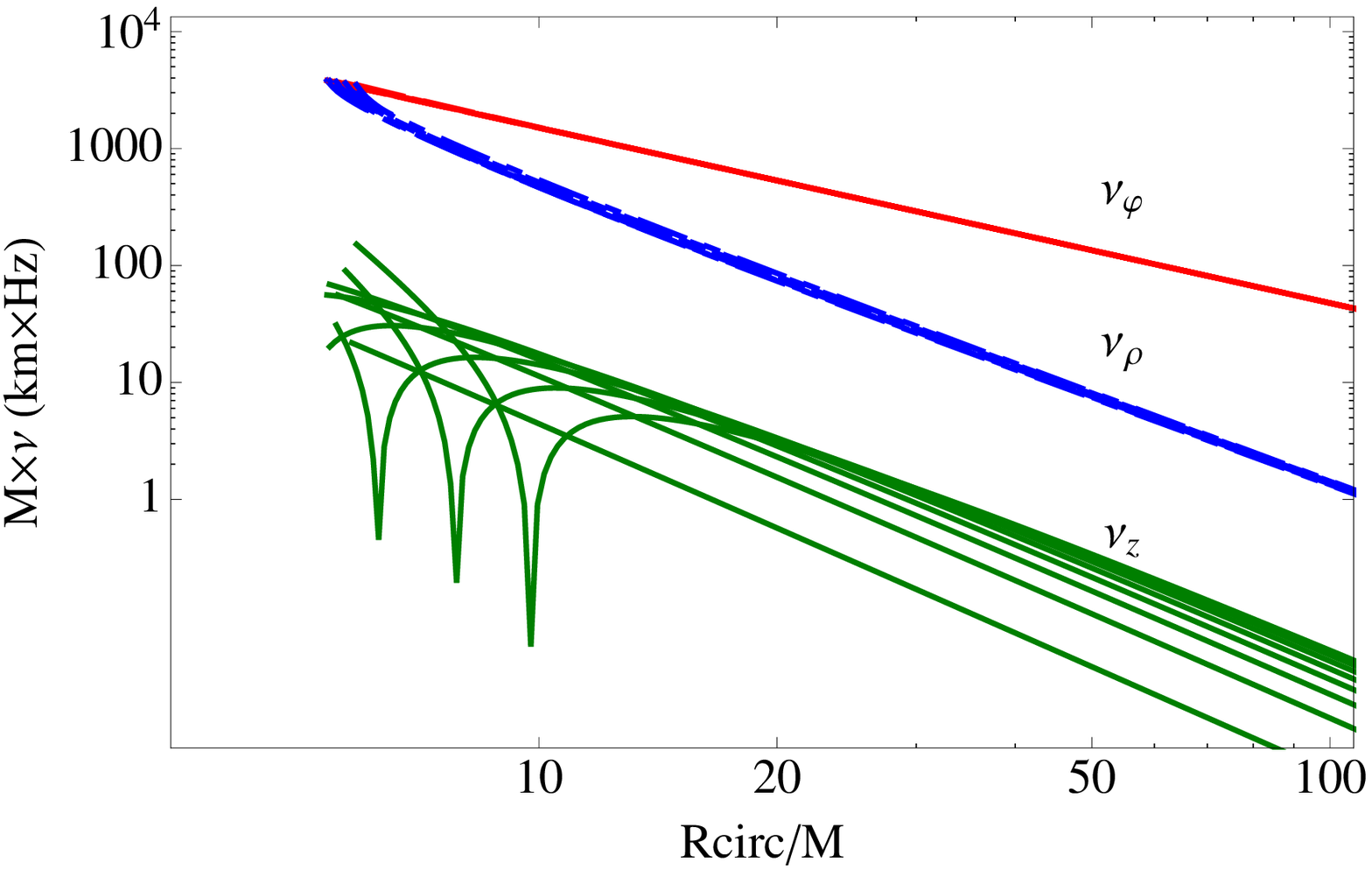} 
\includegraphics[width=.39\textwidth]{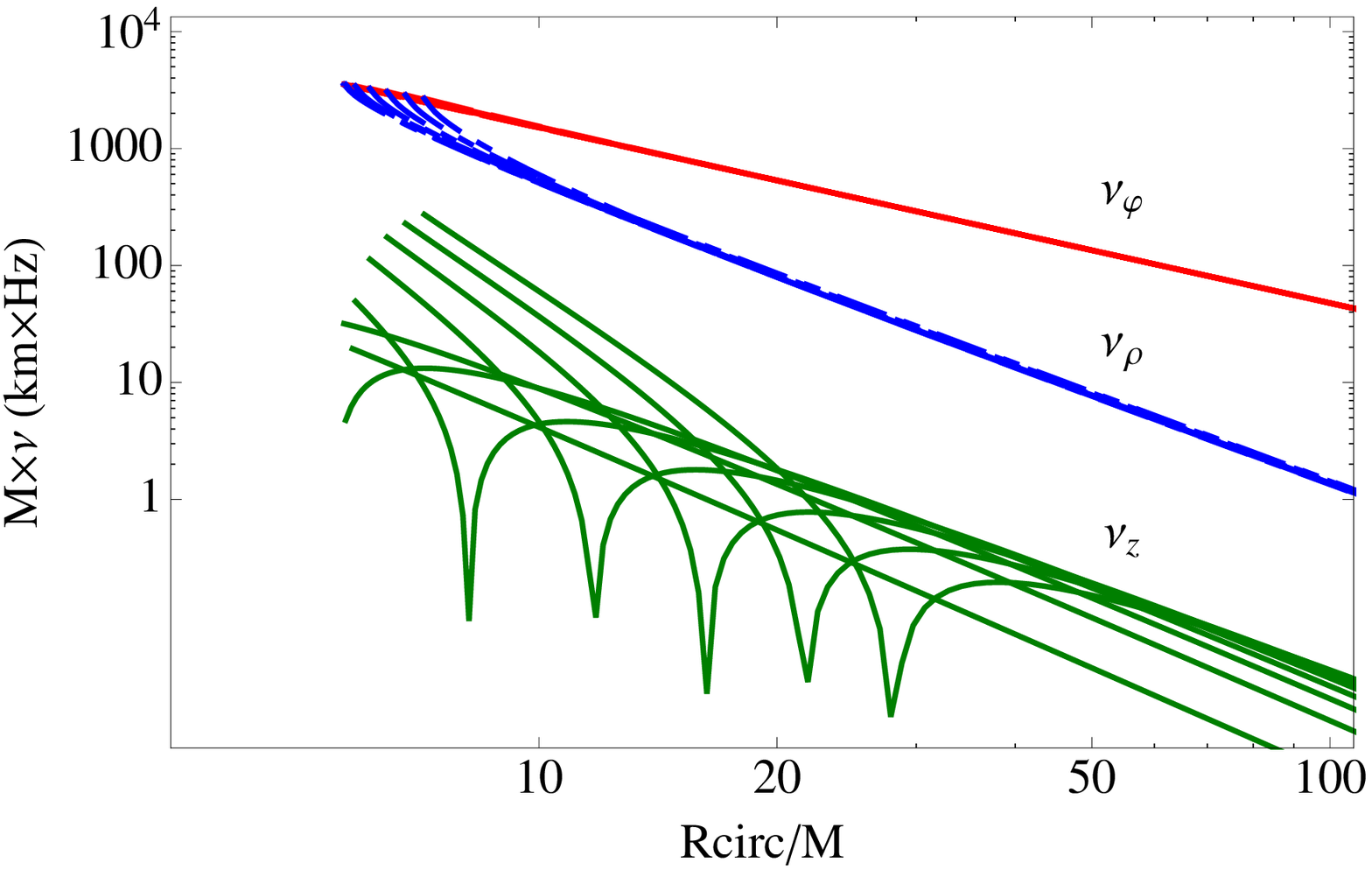}
\includegraphics[width=.39\textwidth]{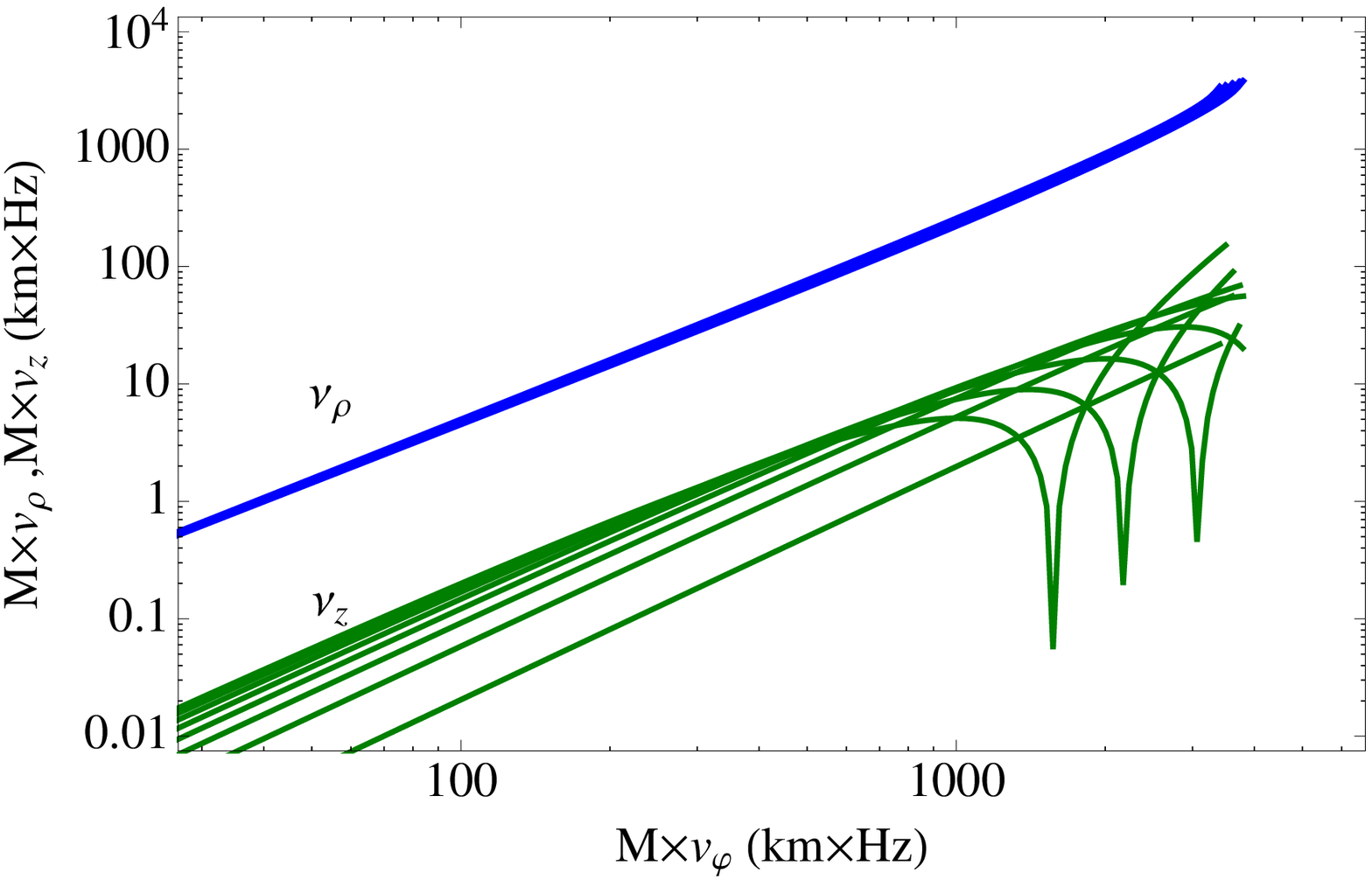} 
\includegraphics[width=.39\textwidth]{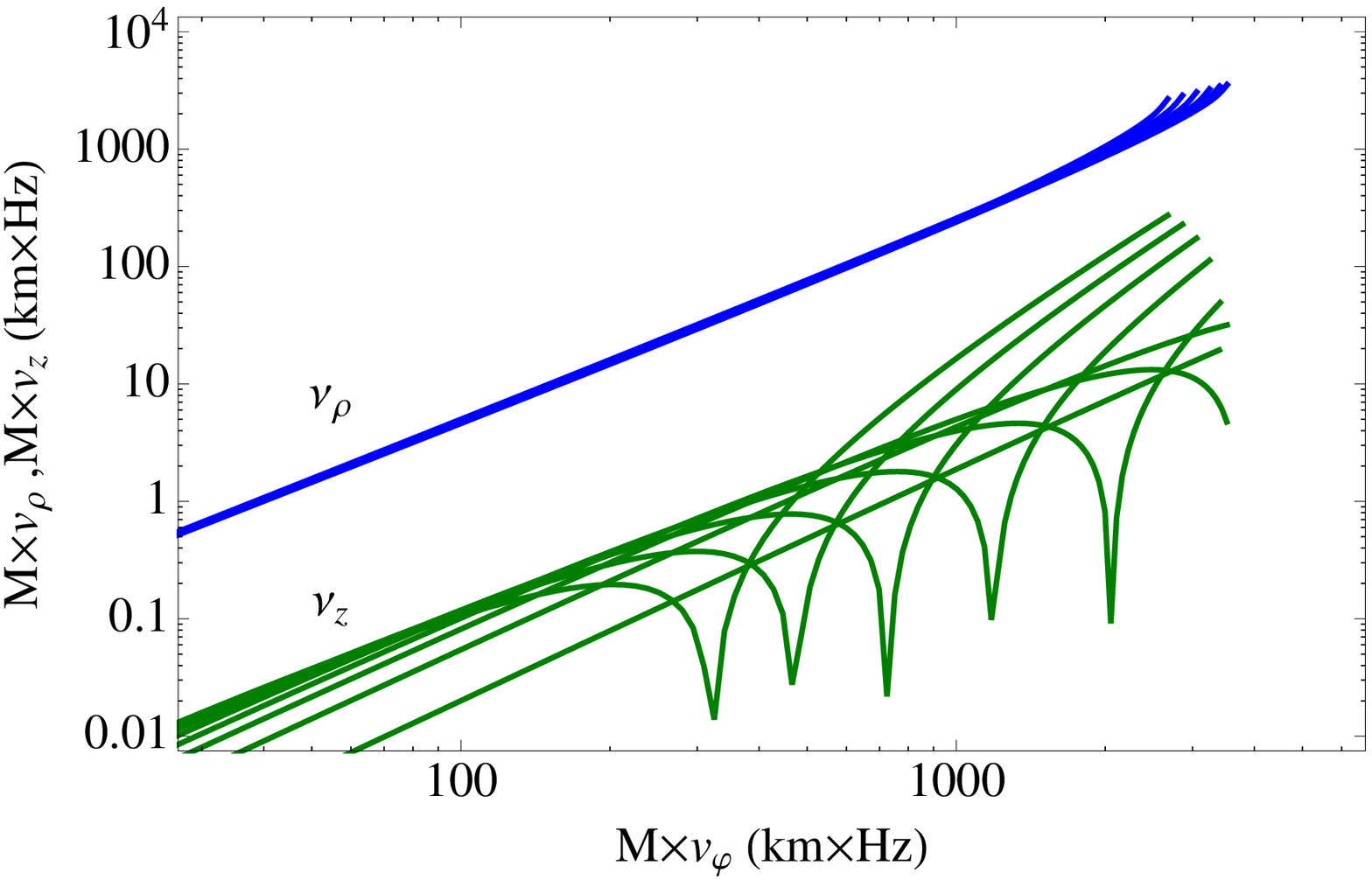}
\caption{Same figure as figure~\ref{Nodalprecession}, with the difference being that the figures in the left column correspond to compact objects with the same value of the quadrupolar deformability $\alpha=5$, while the figures on the right column have $\alpha=9$. The different curves in each figure correspond to different values of the spin parameter in the range $0.05\leq j \leq 0.75$.}
\protect\label{Nodalprecession2}
\end{figure*}

It was found by \cite{PappasQPOs} (and has been recently rediscovered by \cite{Gondeketal,Wisniewiczetal}) that the significant deviation of the quadrupole of a NS from the quadrupole of a Kerr black hole with the same mass and angular momentum, can result in a very different behaviour of the nodal precession frequency from what one would observe for a Kerr spacetime. In particular, if the quadrupolar deformation of the compact object due to rotation is high enough, this can result in a vertical oscillation frequency $\kappa_z$ that is higher than the orbital frequency $\Omega$ in the region close to the ISCO. Then as we move further away from the ISCO, at some point the vertical frequency and the orbital frequency become equal and from that point outward the orbital frequency becomes higher than the vertical frequency. That means that the nodal precession, $\Omega_z=\Omega-\kappa_z$, in the outer region of the spacetime is towards one direction, at the point where the two frequencies are equal there is no precession, and in the inner region the precession is in the opposite direction. 

This effect can be of great significance in modelling accretion discs and would constitute a very clear signature of geodesic precession if it were to be observed in QPOs, as it was concluded by \cite{PappasQPOs}. For the sake of completeness and in order to make the effect more clear so as to facilitate better understanding of the results presented in the main text, we will give here a brief description of the behaviour of the geodesic precession frequencies for different rotation rates and different quadrupolar deformations.

Figs \ref{Nodalprecession} and \ref{Nodalprecession2} present the orbital frequency and the precession frequencies as functions of the distance from the central object (top panels of each figure) as well as the two precession frequencies as functions of the orbital frequency (bottom panels of each figure). All frequencies and distances have been normalized with respect to the mass $M$ of the compact object which provides a natural length scale in geometric units (assumed to be in km). The different frequencies shown here are related to the angular frequencies as $\Omega=2\pi\nu$. The item of interest in all figures are the curves for the nodal precession frequency, $\nu_z=\nu_{\varphi}-\kappa_z/(2\pi)$, while the other frequencies are shown for illustrative purposes and perspective. 

Fig. \ref{Nodalprecession} shows how the precession frequencies behave for two different spin parameters ($j=0.3$ left column, $j=0.4$ right column) if we change the quadrupolar deformation parameter $\alpha$. The values for $\alpha$ are in the range $[1.2,10]$, where the value $\alpha=1$ would correspond to the Kerr quadrupole $M_2=-j^2M^3$. We have chosen these values for the spin parameter because they don't correspond to extremely rapidly rotating NSs. In particular, $j=0.4$ is well within the region where the zero of the nodal precession is outside the surface of the star, as Fig. \ref{OmegaZisco3} shows. We should also note here that lower values of $\alpha$ correspond to NSs with masses that are towards the maximum mass/high density end of the mass-radius diagram, while higher values of $\alpha$ correspond to lower mass NSs (see \cite{poisson,pappas-apostolatos,PappasMoments2,Urbanec2013MNRAS}). 

Returning to the behaviour of the nodal precession, we can see from Fig. \ref{Nodalprecession} that for a given spin parameter, for small values of the deformation parameter, the precession frequency behaves as a power law just like the nodal precession frequency of the Kerr geometry (the leading order behaviour is of the form $\sim r^{-3}$ for Kerr). As $\alpha$ increases though, this power law starts to break at the ISCO (which is at the smallest radius side of the curves, or highest $\nu_{\varphi}$) until for some value of $\alpha$ the first spike appears at a radius outside the ISCO, which corresponds to a zero nodal precession. Beyond that point, for even higher values of $\alpha$ the position of the zero of the nodal precession moves further out until we reach the maximum value of $\alpha$. One should also note that at the far away region (larger radii, smaller orbital frequencies) the curves seem to bundle together, which is what one would expect since at that region the frequencies are mainly characterized by the rotation and the quadrupole or any higher moment should be less important.

Fig. \ref{Nodalprecession2} shows again the various frequencies but in this case we have plotted for two values of the quadrupolar deformability parameter, $\alpha=5$ (left column) and $\alpha=9$ (right column), models with varying spin parameter in the range $j\in[0.05,0.75]$. To give some idea of what sort of NSs that would correspond to, depending on the EoS, these values of $\alpha$ correspond to NSs with masses between $1-1.4M_{\odot}$.  

One can notice (particularly from the far away behaviour of the nodal precession frequencies) that the first effect of increasing rotation is to move upwards in a parallel fashion the frequency curves. This is something that was expected from the behaviour of the Kerr nodal precession frequencies, since the spin parameter multiplies the leading order behaviour of the frequencies ($\sim j r^{-3}$). As in the previous case, we notice that as the rotation increases and with it the magnitude of the quadrupole moment (which is $\propto j^2$), the power law behaviour develops a break close to the ISCO, which then propagates outwards until a zero of the precession frequency appears (in the form of a spike in these log-log plots) which moves outwards, to larger radii or lower orbital frequencies, as the spin parameter continues to increase. The difference between the two columns with the different values of $\alpha$ is that in the case of the higher quadrupolar deformability (right column plots) the position where the precession becomes zero can be further out than in the case of the lower $\alpha$.

Finally for actual NSs, there is also the issue of the location of the surface of the star (presented in Appendix \ref{radii}). It can be for some EoSs that the NS that corresponds to a combination of a spin parameter and a quadrupolar deformation that lead to a vanishing precession frequency at some radius, has actually an equatorial radius that is further out. In these cases, the geodesics, the precession frequencies of which we are investigating, terminate at the surface of the star.    

\bsp \label{lastpage}

\end{document}